\documentclass[notitlepage,twocolumn,letterpaper,natbib,aps,prd,amsmath,amsfonts,nofootinbib,preprintnumbers,superscriptaddress,secnumarabic]{revtex4-2}
\pdfoutput=1
\usepackage{amssymb,amsmath,latexsym,mathrsfs}
\usepackage{url}
\usepackage{enumitem}
\usepackage{graphicx}
\usepackage{booktabs}
\usepackage[usenames,dvipsnames]{color}
\usepackage[breaklinks,colorlinks,urlcolor=blue,citecolor=blue,linkcolor=magenta]{hyperref}
\usepackage{multirow}
\usepackage{float}
\usepackage{cases}
\usepackage{blindtext}
\usepackage{pifont}
\usepackage{hhline}
\usepackage{physics}

\usepackage{xcolor}
\definecolor{linkcolor}{rgb}{0.0, 0.47, 0.75}
\definecolor{citecolor}{rgb}{1.0, 0.5, 0.0}
\hypersetup{
  linkcolor  = linkcolor,
  citecolor  = linkcolor,
  urlcolor   = linkcolor,
  colorlinks = true
}

\begin{document}

\title{On T violation in non-standard neutrino oscillation scenarios}

\author{Thomas Schwetz}%
\affiliation{%
	Institut für Astroteilchenphysik, Karlsruher Institut für Technologie (KIT), 
	76131 Karlsruhe, Germany
}%
\author{Alejandro Segarra}%
\affiliation{%
	Institut für Theoretische Teilchenphysik, Karlsruher Institut für Technologie (KIT), 
	76131 Karlsruhe, Germany
}%

\date{\today}

\begin{abstract}
We discuss time reversal (T) violation in neutrino oscillations in generic new physics scenarios. A general parameterization is adopted to describe flavour evolution, which captures a wide range of new physics effects, including non-standard neutrino interactions, non-unitarity, and sterile neutrinos in a model-independent way. In this framework, we discuss general properties of time reversal in the context of long-baseline neutrino experiments. Special attention is given to fundamental versus environmental T violation in the presence of generic new physics. We point out that T violation in the disappearance channel requires new physics which modifies flavour mixing at neutrino production and detection. We use time-dependent perturbation theory to study the effect of non-constant matter density along the neutrino path, and quantify the effects for the well studied baselines of the DUNE, T2HK, and T2HKK projects. The material presented here provides the phenomenological background for the model-independent test of T violation proposed by us in Ref.~\cite{Schwetz:2021cuj}. 
\end{abstract}

\maketitle

\tableofcontents
\section{Introduction}

The search for CP violation is a central goal for current and upcoming
long-baseline neutrino oscillation experiments. Thanks to the CPT
theorem, fundamental CP violation is closely tied to T violation.
Early works on this topic are
Refs.~\cite{Cabibbo:1977nk,Bilenky:1980cx,Barger:1980jm}. Fundamental
CP or T violation are related to complex couplings in the
Lagrangian. Indeed, in the standard three-flavour scenario
\cite{Zyla:2020zbs}, it is described by a single complex phase in the
lepton mixing
matrix~\cite{Pontecorvo:1967fh,Gribov:1968kq,Maki:1962mu},
the so-called Dirac phase \cite{Kobayashi:1973fv}. However, actual neutrino oscillation
experiments involve neutrino passage through matter, and hence
observable transition probabilities are subject to matter
effects~\cite{Wolfenstein:1977ue}. Since a background of normal matter
leads to violation of CPT in the neutrino flavour evolution
\cite{Langacker:1986jv}, fundamental CP or T violation is not directly
observable, see Refs.~\cite{Bernabeu:2018use,Bernabeu:2019npc} for a
recent discussion.  In this respect, T violation has an advantage over
CP violation for the following reason. Since the matter effect is
different between neutrinos and antineutrinos, environmental CP
violation is typically large and difficult to disentangle from
fundamental CP violation. In contrast, it is well known that a
symmetric matter density profile (symmetric between neutrino source
and detector) does not introduce environmental T violation if the
fundamental theory is T invariant, see for instance
Refs.~\cite{Krastev:1988yu,Akhmedov:2001kd}. On the other hand, T
violation itself is again difficult to observe experimentally, since
it formally corresponds to exchanging neutrino flavours of neutrino
source and detector. There is extensive literature on T violation in
neutrino oscillations, see
Refs.~\cite{Cabibbo:1977nk,Kuo:1987km,Krastev:1988yu,Toshev:1989vz,Toshev:1991ku,Arafune:1996bt,Parke:2000hu,Akhmedov:2001kd,Xing:2013uxa,Petcov:2018zka,Bernabeu:2019npc}
for an incomplete list.

The usual search for CP violation is highly model-dependent. It
relies on the standard unitary three-flavour paradigm, implying the
absence of any new physics in neutrino interactions, mixing, and
propagation. In this restricted framework, a parametric fit to the
available data is performed in terms of the standard mass-squared
differences $\Delta m^2_{21}$, $\Delta m^2_{31}$, mixing angles
$\theta_{12},\theta_{23},\theta_{13}$, and the complex phase
$\delta$. ``Discovery of CP violation'' is usually identified with the
situation when the fit disfavours values of $\delta = 0$ and $\pi$ at
a certain confidence level. Indeed, within this approach, current
results from the T2K~\cite{Abe:2019vii} and NOvA~\cite{Acero:2019ksn}
long-baseline experiments, combined with the global neutrino
oscillation data, show already an indication for a preferred range
of $\delta$ \cite{Esteban:2020cvm,deSalas:2020pgw,Capozzi:2021fjo}.

In Ref.~\cite{Schwetz:2021cuj}, we have proposed a method, with the
goal to address several limitations outlined above and to search for
fundamental T violation in the neutrino sector in a more
model-independent way. In order to achieve this goal, we have
introduced two main ingredients:
\begin{enumerate}
\item[($i$)] We prosed a rather general
parameterization of neutrino evolution, to describe the flavour system
more model-independently, and
\item[($ii$)] we presented a
potentially realistic way to search for fundamental T violation in
long-baseline experiments.
\end{enumerate}
Regarding ($i$), our general parameterization allows for effects of
non-standard interactions in neutrino source and detection as well as
arbitrary matter effect. Mixing can be non-unitary, and therefore the
presence of sterile neutrinos is allowed, as long as they do not
introduce additional oscillation frequencies (i.e., we restric to two
independent oscillation frequencies). We will review this
parameterization in detail in sec.~\ref{sec:formalism} below.

The main idea with respect to item ($ii$) is the following: we
consider the oscillation probabilities within the general framework at
different baselines but at the same energy. The reason for this
assumption is that we want to be agnostic about the energy dependence
of the new physics. Hence we need to combine data from long-baseline
experiments at different baselines $L$ at the same neutrino
energy. Then we check if the data requires T-odd (or equivalently
$L$-odd) terms in the transition probability, based on the
model-independent parameterization.
We have shown in Ref.~\cite{Schwetz:2021cuj} that this test
potentially can be performed already with data from 3 different
long-baseline experiments (plus data from a near detector). This opens
the possibility to apply the proposed test with actually planned and
proposed experiments, such as DUNE ($L=1300$~km)
\cite{Abi:2020wmh,Abi:2020evt}, T2HK ($L=295$~km) \cite{Abe:2018uyc},
T2HK with a second detector in Korea, T2HKK ($L=1100$~km)
\cite{Abe:2016ero}, and a long-baseline experiment at the European
Spalation Source, ESS$\nu$SB
($L=540$~km)~\cite{Baussan:2013zcy,Blennow:2019bvl}. The crucial
requirement is the availability of measurments at the first and the
second oscillation maxima (at the same energy), with sufficiently good
energy reconstruction. Preliminary sensitivity estimates have been
performed in Ref.~\cite{Schwetz:2021cuj}.

The goals of the present paper are the following. We provide a more
in-depth discussion of T violation, allowing for the general
non-standard physics described above to establish the theoretical
basis for the test proposed in Ref.~\cite{Schwetz:2021cuj}. We show
explicitly that several results known for the standard framework carry
over to the new-physics case considered here. For instance, we prove
that any non-standard matter effect does not introduce environmental T
violation if the fundamental theory is T conserving, as long as the
matter density profile is symmetric. Special care is given to
non-standard mixing effects in source and detector. We give a careful
definition of the time reversal symmetry and discuss its effect in
non-standard mixing scenarios. Along this way we establish the basic assumptions
of the test in~\cite{Schwetz:2021cuj}.

In Ref.~\cite{Schwetz:2021cuj}, we have formulated the test by
assuming a constant matter density along the neutrino path, and that
the matter density is the same for all experiments. These assumptions are
only approximately valid for the experiments under consideration.
Therefore, in the present article we provide a quantitative estimate
for corrections induced by realistic matter density profiles, based on
the detailed investigations for the T2HKK and DUNE baselines from
Refs.~\cite{Hagiwara:2011kw, Roe:2017zdw}. In general, our approach is
based on a perturbation ansatz, using that both the new physics as
well as the non-constant density effects are small perturbations to
the standard three-flavour and constant matter case.  The methods
developed below allow for a straightforward correction of the T
violation test with respect to non-constant density.

The outline of the remainder of the paper is as follows.  In
section~\ref{sec:formalism} we review the model-independent
parameterization of the neutrino flavour evolution and derive the
transition amplitudes and probabilities, taking into account
non-constant matter densities as a small perturbation effect. In
section~\ref{sec:T} we consider the time reversal transformation
within this model-independent setting and discuss which properties
known for the standard oscillation case still hold in our
model-independent framework and which not.  In section~\ref{sec:disapp}
we provide some further discussion and comment on T violation in the
disappearance channel within new physics scenarios.  In
section~\ref{sec:corrections} we provide quantitative estimates of
non-constant matter density profiles for T2HK, T2HKK and DUNE. We conclude
in section~\ref{sec:conclusion}.

\section{Model-independent description of flavour evolution}
\label{sec:formalism}

We assume that propagation of the three SM neutrino states is
described by a hermitian Hamiltonian $H(E,x)$, which depends on
neutrino energy $E$ and in general on the matter density at the
position $x$ along the neutrino path. We follow the usual
approximation in describing neutrino evolution by setting, $x=t$ (in
units where the speed of light is unity), i.e., localized neutrino
wave packets propagating with the speed of light. Therefore, the space
dependence of the Hamiltonian effectively becomes a time
dependence. In the following we will use $x$ and $t$ interchangeably.
The evolution of the flavour state $|\psi\rangle$ is described by the equation
\begin{equation}\label{eq:schroedinger}
  i\partial_t|\psi\rangle = H(t) |\psi\rangle \,,
\end{equation}
where here and in the following we suppress the energy dependence.

We consider neutrinos propagating from a source at position $x_s$ at
time $t_s$ to a detector at position $x_d$, arriving at time $t_d$,
with $x_d-x_s = t_d - t_s = L$. Let us define
\begin{align}
  H(t) &= H_{\rm vac} + V_{\rm tot}(t) = H_0 + V(t) \,, \\
  V_{\rm tot}(t) &= V_0 + V(t) \,,\\
  H_0 &= H_{\rm vac} + V_0 \,,
  \label{eq:H0}
\end{align}
where $H_0$ contains contributions from the vacuum Hamiltonian as well
as the average matter potential $V_0$ and is time/position independnet,
whereas $V(t)$ corresponds to the matter potential due to the varying
matter density, with $\int_{x_s}^{x_d} dx V(x) = 0$. For our purposes
(long-baseline experiments) we assume that $V(t)$ is a small
perturbation, i.e., that the matter density is roughly constant along
the neutrino path. This is a good approximation for
experiments with baselines less than several
1000~km~\cite{Miura:2001pi,Yokomakura:2002av,Hagiwara:2011kw,Roe:2017zdw}.
We will quantify this in section~\ref{sec:corrections}.

Let us diagonalize the position-independent part $H_0$ by $H_0 = W
\lambda W^\dagger$, with $W$ being a unitary matrix and $\lambda =
(\lambda_i)$ a diagonal matrix of the real eigenvalues
$\lambda_1,\lambda_2,\lambda_3$ of $H$. Both $W$ and $\lambda$ depend
on the neutrino energy and they will be different for neutrinos and
antineutrinos due to the matter effect as well as fundamental CP violation.
The energy eigenstates $|\nu_i\rangle$ fulfill
\begin{align}\label{eq:H0EV}
  H_0 |\nu_i\rangle = \lambda_i |\nu_i\rangle \,.
\end{align}

We allow for arbitrary (non-unitary) mixing of the energy
eigenstates $|\nu_i\rangle$ with the flavour states $|\nu_\alpha\rangle$ relevant for
detection and production,
\begin{equation}\label{eq:mixing}
  |\nu_\alpha^{s,d}\rangle = \sum_{i=1}^3 (N_{\alpha i}^{s,d})^* |\nu_i\rangle \,,
\end{equation}
where $^*$ denotes complex conjugation.  We make no specific
assumption on the coefficients $N_{\alpha i}^{s}$ and $N_{\alpha
  i}^{d}$. They can include effects of heavy sterile neutrinos as well
as non-standard interactions.  Note that generically the unitary
matrix $W$ diagonalizing the Hamiltonian will contribute to these
coefficients; in specific models they may be related and the new
physics entering in the coefficients $N_{\alpha i}^{s,d}$ and will also
induce non-standard contributions to the Hamiltonian $H(t)$. Some
examples for specific models are non-unitary
mixing~\cite{FernandezMartinez:2007ms,Escrihuela:2016ube},
non-standard neutrino
interactions~\cite{Ge:2016dlx,deGouvea:2015ndi,Coloma:2019mbs,Denton:2020uda},
or the presence of sterile
neutrinos~\cite{Kopp:2013vaa,Gandhi:2015xza,Palazzo:2015gja,Berryman:2015nua}.
Our oscillation formalism has some similarities with the one developed
in the context of non-unitary mixing, see e.g.,
Refs.~\cite{Antusch:2006vwa,Escrihuela:2015wra,Fong:2017gke}.
Ref.~\cite{Blennow:2016jkn} discusses the parametric relation between
various non-standard scenarios.

Here we want to be more general and treat $W$ and
$N^{s,d}$ as independent. In particular, $N^{s,d}$ can be non-unitary,
arbitrary functions of neutrino energy, and they can be different for
processes relevant at the neutrino source and for detection (as
indicated by the indicies $s$ and $d$). This implies that in general
$|\nu_\alpha^{s}\rangle \neq |\nu_\alpha^{d}\rangle$. But we do assume
that $N_{\alpha i}^{s,d}$ are the same for different experiments (at
the same energy). Note, however, that while the mixing in
Eq.~\eqref{eq:mixing} can be non-unitary, the induced matter potential
$V_{\rm tot}$ as well as the total Hamiltonian will be still
hermitian, leading to a unitary evolution of the system via
Eq.~\eqref{eq:schroedinger}.\footnote{Per assumption we do not allow
  for the decay of energy eigenstates, which would lead to a
  non-unitary evolution equation.}

As we will see in the following, complex phases of $N_{\alpha i}^{s,d}$
induce fundamental CP and T violation. In the standard scenario there
is only one relevant phase (the Dirac CP phase), whereas in
non-standard scenarios there are several new sources for complex
phases~\cite{FernandezMartinez:2007ms,Escrihuela:2016ube,Ge:2016dlx,deGouvea:2015ndi,Coloma:2019mbs,Denton:2020uda,Kopp:2013vaa,Gandhi:2015xza,Palazzo:2015gja,Berryman:2015nua}.

\subsection{Transition probabilities}

We are interested in the transition amplitude for a neutrino of
flavour $\alpha$ at the source to a neutrino of flavour $\beta$ at the
detector, $\mathcal{A}(\nu_\alpha^s\to\nu_\beta^d) \equiv \mathcal{A}_{\alpha\beta}$,
and the corresponding transition probability
$P_{\alpha\beta} = |\mathcal{A}(\nu_\alpha^s\to\nu_\beta^d)|^2$. Consider first
the unitary evolution operator $S$ of the energy eigenstates $|\nu_i(t^s)\rangle
\to |\nu_j(t^d)\rangle$: $S_{ij}(t_d,t_s)$. Then using Eq.~\eqref{eq:mixing} we obtain 
\begin{align}\label{eq:ampl-general}
  \mathcal{A}_{\alpha\beta} &=
  \langle \nu_\beta^d | S(t_d,t_s)|\nu_\alpha^s\rangle 
  = \sum_{ij} S_{ij}(t_d,t_s) N_{\alpha i}^{s*} N_{\beta j}^d \,.
\end{align}

Let us now consider $V(t)$ as a small perturbation and solve
Eq.~\eqref{eq:schroedinger} using time-dependent perturbation theory,
see e.g. Ref.~\cite{sakurai}.
At zeroth order in $V(t)$, i.e., assuming only the
constant Hamiltonian $H_0$, the evolution operator is just given by
$S_{ij}^{(0)}(t_d,t_s) = e^{-i\lambda_i(t_d-t_s)}\delta_{ij}$ and
\begin{equation}\label{eq:A0}
  \mathcal{A}_{\alpha\beta}^{(0)} = \sum_i N_{\alpha i}^{s*} N_{\beta i}^d e^{-i\lambda_i(t_d-t_s)} \,. 
\end{equation}
At first order in $V(t)$ we find
\begin{align}
  &S_{ij}^{(1)}(t_d,t_s) = -i e^{-i\lambda_j t_d + i\lambda_i t_s} 
  \int_{t_s}^{t_d} dt V_{ij}(t) e^{i(\lambda_j-\lambda_i)t}  \,, \\
  &\mathcal{A}_{\alpha\beta}^{(1)} =
  -i \sum_{ij} N_{\alpha i}^{s*} N_{\beta j}^d e^{-i\lambda_j t_d + i\lambda_i t_s}
  \int_{t_s}^{t_d} dt V_{ij}(t) e^{i(\lambda_j-\lambda_i)t} \,. \label{eq:A1}
\end{align}
For the transition probability we have
\begin{align}
P_{\alpha\beta} & \approx \left| \mathcal{A}_{\alpha\beta}^{(0)} + \mathcal{A}_{\alpha\beta}^{(1)}
\right|^2 \approx P_{\alpha\beta}^{(0)} + P_{\alpha\beta}^{(1)} \,, \nonumber\\
P_{\alpha\beta}^{(0)} &= \left| \mathcal{A}_{\alpha\beta}^{(0)} \right|^2 \,,\quad
P_{\alpha\beta}^{(1)} = 2 \Re \left[ {\mathcal{A}_{\alpha\beta}^{(0)}}^* \mathcal{A}_{\alpha\beta}^{(1)} \right] \,.  
	\label{eq:Pcorr}
\end{align}  

The parameterization discussed here allows to cover a rather broad range of new physics scenarios, including non-standard interactions in charged current and neutral current interactions, generic non-unitarity, as well as sterile neutrinos (as long as they do not introduce additional oscillation frequencies for the relevant energy and baselines). In principle we can allow for an arbitrary energy dependence of $N_{\alpha i}^{s,d}$ and $\lambda_i$, although in the presence of finite energy resolution we have to demand that the energy dependence is weak at the scale of the resolution. Following Ref.~\cite{Schwetz:2021cuj}, due to the success of the standard three-flavour paradigm, we can assume that new physics effects are a small perturbation to the standard case. To leading order, our parameterization covers also non-standard interactions within the most general effective field theory framework \cite{Falkowski:2019kfn,Bischer:2019ttk}.\footnote{We thank Martin Gonzalez-Alonso for illuminating discussions about this point.}

\section{Properties under the T transformation}
\label{sec:T}

With these generalized expressions for the transition amplitudes and
probabilities at hand, we can now discuss their properties under the
time reflection transformation.

\subsection{Constant matter potential}
\label{sec:T_const}

Let us first assume that the matter potential is constant, i.e., we work at zeroth order in the time-dependent perturbation theory. In this case we can define the T transformation by 
\begin{equation}\label{eq:T}
  {\rm T:}\, t\to -t \,.
\end{equation}
Then we obtain from Eq.~\eqref{eq:A0}:
\begin{equation}\label{eq:TA}
{\rm T} \, \mathcal{A}_{\alpha\beta}^{(0)} = {\rm T} \,\mathcal{A}^{(0)}(\nu_\alpha^s\to\nu_\beta^d) 
= \left[\mathcal{A}^{(0)}(\nu_\beta^d \to \nu_\alpha^s)\right]^* \,. 
\end{equation}
Note that if $N_{\alpha i}^s \neq N_{\alpha i}^d$ then
$\mathcal{A}^{(0)}(\nu_\beta^d \to \nu_\alpha^s) \neq
\mathcal{A}_{\beta\alpha}^{(0)} = \mathcal{A}^{(0)}(\nu_\beta^s \to
\nu_\alpha^d)$.  Therefore, the usual result T$P_{\alpha\beta} =
P_{\beta\alpha}$ holds only if the mixing is the same for the
processes relevant for neutrino production and detection. If there is
new physics distinguishing between neutrino production and detection,
the transformation \eqref{eq:T} is \emph{not} equivalent to exchanging
only the neutrino \emph{flavours} of source and detector, but also the type of
interaction needs to be exchanged (formally $d\leftrightarrow s$).

For real $N_{\alpha i}^{s,d}$, it follows from Eq.~\eqref{eq:A0} that
T$\mathcal{A}_{\alpha\beta}^{(0)} = \mathcal{A}_{\alpha\beta}^{(0)*}$ and
therefore T$P_{\alpha\beta} = P_{\alpha\beta}$. As expected, the T
transformation tests complex phases in the theory.

For the transition probabilities at zeroth order in $V(t)$ we obtain
\begin{align}
  P_{\alpha\beta}^{(0)} =& \left | \sum_{i} c_i e^{-i\lambda_i(t_d-t_s)} \right|^2 
  \\
  =& \sum_i |c_i|^2 + 2\sum_{j<i} \Re(c_i c_j^{*})\cos(\omega_{ij}L)
  \nonumber\\
  &-2\sum_{j<i} \Im(c_i c_j^{*})\sin(\omega_{ij}L) \,, \label{eq:prob-gen}
\end{align}
with the abbreviation $c_i \equiv N_{\alpha i}^{s*} N^{d}_{\beta
  i}$ and the frequencies $\omega_{ij} \equiv \lambda_j-\lambda_i$. As
usual we identify the baseline by $L=t_d-t_s$.  Therefore, T 
is formally equivalent to $L \to -L$. We see that the first line of
Eq.~\eqref{eq:prob-gen} is invariant under T, whereas the
second line is T-odd.  It is also apparent that T violation will be
present only for non-zero $\Im(c_i c_j^{*})$, i.e., non-trivial
complex phases of $N_{\alpha i}^{s,d}$.

Hence, fundamental T violation can be established by proving the
presence of the $L$-odd term in the probability. Or, put in other
words, if data cannot be described by an $L$-even transition
probability
\begin{equation}
	P_{\alpha\beta}^\mathrm{even} (L) =
	\sum_{i} c_i^2 
	+2 \sum_{j<i} c_i c_j \cos(\omega_{ij} L) 
	\label{eq:ProbEven}
\end{equation}
with $c_i$ real, fundamental T violation needs to be present in the
theory. This is the test proposed in Ref.~\cite{Schwetz:2021cuj}.

\subsection{Non-constant matter potential}
\label{sec:T_nonconst}

Let us now consider the first-order correction in the case of a
non-constant matter potential $V(t)$. We recall that $V(t)$ is defined between the locations of the source $x_s$ and the detector $x_d$. Therefore, we have to make sure that $V(t)$ is evaluated only for times in the interval $[t_s, t_d]$. This requirement has to be respected also when applying the T transformation in Eq.~\eqref{eq:T}. One possible choice is to replace Eq.~\eqref{eq:T} by
\begin{equation}\label{eq:Talt}
  {\rm T:} \, t \to t_s+t_d-t \,.
\end{equation}
This implies that T leads to $t_s \leftrightarrow t_d$ and that T is still equivalent to $L\to -L$. 
Applying this transformation to Eq.~\eqref{eq:A1} we find
\begin{equation}\label{eq:TA1}
  {\rm T}\mathcal{A}_{\alpha\beta}^{(1)} =
  i \sum_{ij} N_{\alpha j}^{s*} N_{\beta i}^d e^{i\lambda_j t_d - i\lambda_i t_s}
  \int_{t_s}^{t_d} dt V_{ij}^*(t) e^{-i(\lambda_j-\lambda_i)t} \,.  
\end{equation}
Here we have re-named the indices $i\leftrightarrow j$ and used the
hermiticity of the Hamiltonian, $V_{ij} = V_{ji}^*$. We observe 
that Eq.~\eqref{eq:TA}, which we have obtained for
$\mathcal{A}_{\alpha\beta}^{(0)}$, as well as the comments thereafter
hold also for the first-order correction
$\mathcal{A}_{\alpha\beta}^{(1)}$.

As we have seen in sec.~\ref{sec:T_const} for the case of constant
matter density, the mixing coefficients $N_{\alpha i}^{s,d}$ are the
only sources of complex phases relevant for the transition
probabilities and therefore fundamental T violation/conservation can
be characterized by the presence/absence of (non-trivial) complex
phases in $N_{\alpha i}^{s,d}$. Note, however,
that $N_{\alpha i}^{s,d}$ are defined with respect to the eigenbasis
of the Hamiltonian $H_0$ for constant density, see Eqs.~\eqref{eq:H0EV}
and \eqref{eq:mixing}. Therefore, in general the non-constant
perturbation $V(t)$ may contain non-trivial complex phases even if
$N_{\alpha i}^{s,d}$ are real. Hence, we will define ``fundamental T
conservation'' in the following as real $N_{\alpha i}^{s,d}$
\emph{and} real $V_{ij}(t)$.\footnote{Note that $N_{\alpha i}^{s,d}$
  includes possible phases from the constant matter
  potential. Therefore, having complex phases in $V(t)$ but not in in
  $N_{\alpha i}^{s,d}$ would require rather special CP violating new
  physics, coupling only to density \emph{variations}.}

Comparing Eqs.~\eqref{eq:A1} and \eqref{eq:TA1}, we see that even for
$N_{\alpha i}^{s,d}$ and $V_{ij}(t)$ real, we still have
T$\mathcal{A}_{\alpha\beta}^{(1)} \neq
\mathcal{A}_{\alpha\beta}^{(1)*}$ in general. Hence, we recover the
result that a non-constant matter density induces environmental T
violation in neutrino oscillations, even if the fundamental theory is
T conserving. This is a well-known result in the standard oscillation
scenario, e.g., \cite{Akhmedov:2001kd}.

We can also prove the result known for the standard case, namely that
a symmetric matter profile does not induce T violation in the absence
of fundamental T violation.  Following Ref.~\cite{Akhmedov:2001kd},
this can most easily seen by setting the time $t=0$ at the baseline
mid-point, such that the time interval $[t_s,t_d]$ becomes symmetric:
$[-L/2,L/2]$. The transformation~\eqref{eq:Talt} still remains $L
\to -L$. Rewriting Eq.~\eqref{eq:A1} we obtain
  \begin{widetext}
  \begin{align}
	  \label{eq:A(1)}
  &\mathcal{A}_{\alpha\beta}^{(1)} =
  -i \sum_{ij} N_{\alpha i}^{s*} N_{\beta j}^d e^{-i(\lambda_j + \lambda_i)L/2}
  \int_{-L/2}^{L/2} dt V_{ij}(t) e^{i(\lambda_j-\lambda_i)t} \,, \\
	  \label{eq:TA(1)}
  &{\rm T}\mathcal{A}_{\alpha\beta}^{(1)} =
  i \sum_{ij} N_{\alpha i}^{s*} N_{\beta j}^d e^{i(\lambda_j + \lambda_i)L/2}
  \int_{-L/2}^{L/2} dt V_{ij}(-t) e^{-i(\lambda_j-\lambda_i)t} \,,    
  \end{align}    
  \end{widetext}
where in the second line we have performed the variable transformation
$t\to-t$ in the integral. Comparing the two expressions, we see that
if $N_{\alpha i}^{s,d}$ and $V_{ij}$ are real (no fundamental T
violation) and the matter potential is symmetric, $V_{ij}(t) =
V_{ij}(-t)$, then T$\mathcal{A}_{\alpha\beta}^{(1)} =
\mathcal{A}_{\alpha\beta}^{(1)*}$. Since the same holds for
$\mathcal{A}_{\alpha\beta}^{(0)}$, we have T$P_{\alpha\beta} =
P_{\alpha\beta}$ in this case. Hence, the above statement is
proven. As in the standard scenario,
in order to introduce environmental T violation in the absence of
fundamental T violation an asymmetric matter potential is needed.

\bigskip

Here we derived this result at first order in perturbation theory. In
Ref.~\cite{Akhmedov:2001kd} this statement was proven for the standard
oscillation scenario for arbitrary matter profile. It is
straight-forward to generalize the prove given
in~\cite{Akhmedov:2001kd} also to the non-standard scenario considered
here, which we briefly outline in the following. We depart from Eq.~\eqref{eq:ampl-general} and use general properties of the evolution operator:
\begin{align}
  S(t_d,t_s) S(t_s,t_d) = 1\,,\quad
  S(t_d,t_s) S^\dagger(t_d,t_s) = 1 \,.
\end{align}
The second relation follows from the unitarity of the evolution due to
the hermiticity of $H(t)$. Using now the T transformation T: $t_s\leftrightarrow t_d$ and combining the two properties above we find
\begin{align}\label{eq:TS}
  {\rm T} S(t_d,t_s) = S(t_s,t_d) = S^\dagger(t_d,t_s) \,.
\end{align}

Using again a symmetric time coordinate, we consider the evolution operator $S(t,-t)$. 
Its time evolution is given by~\cite{Akhmedov:2001kd}:
\begin{align}
  i\frac{d}{dt}S(t,-t) = H(t)S(t,-t) + S(t,-t)H(-t) \,.
\end{align}
Take now the transpose of this equation. If the Hamilton operator is real (no fundamental T violation), then it is symmetric. If in addition the density profile is symmetric, $H(t) = H(-t)$, we see that $S(t,-t)$ and $S^T(t,-t)$ follow the same evolution equation and are therefore equal, i.e., $S$ is symmetric. Using this in Eq.~\eqref{eq:TS}, we obtain
T$S(t_d,t_s) = S^*(t_d,t_s)$. With real $N^{s,d}_{\alpha i}$ we obtain then from Eq.~\eqref{eq:ampl-general} that  T$\mathcal{A}_{\alpha\beta} = \mathcal{A}_{\alpha\beta}^*$ and therefore, 
T$P_{\alpha\beta} = P_{\alpha\beta}$.

\section{Comments on T asymmetries and the disappearance channel}
\label{sec:disapp}

Let us collect a few relations regarding time reversal asymmetries. We
define the T asymmetry as
\begin{equation}\label{eq:asym}
	A_{\alpha\beta} = P_{\alpha\beta} - {\rm T} P_{\alpha\beta} \,.
\end{equation}
From Eq.~\eqref{eq:prob-gen} we find for the  zeroth-order asymmetry
\begin{equation}
  A^{(0)}_{\alpha\beta} = 4 \sum_{i<j}
  \Im\left( N_{\alpha i}^s N_{\beta i}^{d*}N_{\alpha j}^{s*} N_{\beta j}^{d} \right)
  \sin\left(\omega_{ij}L\right)\,. \label{eq:A0dis}
\end{equation}
The 1st order correction to the probabilities from
Eq.~\eqref{eq:Pcorr} can be written in the following way:
\begin{align}
	\nonumber
 P_{\alpha\beta}^{(1)} 
	=& \sum_{ijk} \sin(\lambda_{ij}^k \frac{L}{2}) \Re G_{ij}^k	\\
	 &+ \sum_{ijk} \cos(\lambda_{ij}^k \frac{L}{2}) \Im G_{ij}^k \,,
	 \label{eq:A0A1}
\end{align}
where we have defined 
\begin{align}
\lambda_{ij}^k & = 2\lambda_k - (\lambda_i + \lambda_j) \,,\\
G_{ij}^k &= 2 \, N_{\alpha k}^s N_{\beta k}^{d*} N_{\alpha i}^{s*} N_{\beta j}^d \, I_{ij} \label{eq:G}\\
	I_{ij} &= \int_{-L/2}^{L/2} dt V_{ij}(t) e^{i(\lambda_j-\lambda_i)t} \,, 
\end{align}
with $\lambda_{ij}^k = \lambda_{ji}^k$ and $I_{ji} = I_{ij}^*$. The T
transformation corresponds to $L\to - L$ and it follows that T$I =
-I$, T$G = -G$, and therefore the first (second) line in
Eq.~\eqref{eq:A0A1} is T even (odd). Hence, we obtain for the 1st
order asymmetry
\begin{align}\label{eq:A1dis}
  A_{\alpha\beta}^{(1)} = 2 \sum_{ijk} \cos(\lambda_{ij}^k \frac{L}{2}) \Im G_{ij}^k \,.
\end{align}
For a symmetric profile $V_{ij}(t) = V_{ij}(-t)$, we have
\begin{equation}
  I_{ij}^{\rm sym} = \int_{-L/2}^{L/2} dt V_{ij}(t) \cos\left(\omega_{ij} t\right) \,.
\end{equation}
In the absence of fundamental T violation, with $N_{\alpha i}^{s,d}$
and $V_{ij}$ real, $I_{ij}^{\rm sym}$ and $G_{ij}^k$ are real as well,
and $A_{\alpha\beta}^{(1)}=0$, in agreement with our results in
sec.~\ref{sec:T_nonconst}.

\bigskip

Consider now the disappearance probabilities $\beta=\alpha$. In the
standard oscillation scenario, the T transformation becomes trivial,
since exchanging initial and final flavour has no effect. Let us
re-consider this case in the extended new physics scenario.  First we
assume that mixing is identical at source and detector $N_{\alpha i}^s
= N_{\alpha i}^d$. Then we find that both, $A^{(0)}_{\alpha\alpha} =
0$ and $A^{(1)}_{\alpha\alpha} = 0$. The first follows directly from
Eq.~\eqref{eq:A0dis}. The second follows from Eq.~\eqref{eq:A1dis} by noting
that for $N_{\alpha i}^s = N_{\alpha i}^d$ and $\alpha=\beta$ we have
$G_{ij}^k = G_{ji}^{k*}$. Hence, we conclude that for $N_{\alpha i}^s
= N_{\alpha i}^d$ (which includes also standard mixing) no T asymmetry
can be observed in the disappearance channel. This holds even in
presence of complex phases as well as asymmetric matter density
profiles.

On the contrary, if $N_{\alpha i}^s \neq N_{\alpha i}^d$, in the
presence of non-trivial complex phases we obtain
$A^{(0)}_{\alpha\alpha} \neq 0$. If $N_{\alpha i}^s \neq N_{\alpha
  i}^d$ but both real, then $A^{(1)}_{\alpha\alpha} \neq 0$ for an
asymmetric density profile.\footnote{Note, however, that this would be
  a second order effect, being suppressed by the small density
  variations \emph{and} the new physics respondsible for $N_{\alpha
    i}^s \neq N_{\alpha i}^d$.} We conclude that  
\begin{itemize}
\item the observation of T violation in a disappearance channel would
  be a signal of new physics inducing different flavour mixing at
  source and detector;
\item if effects of asymmetric matter densities can be neglected, it
  requires fundamental T violation (in addition to $N_{\alpha i}^s
  \neq N_{\alpha i}^d$).
\end{itemize}

Let us briefly comment on the possible observability of such an
effect, at least in principle. One can follow the approach of
Ref.~\cite{Schwetz:2021cuj} and imagine measurments of
$P_{\alpha\alpha}(L_b)$ at a number of baselines $L_b$ at a fixed
energy, and in this way study the $L$ dependence of the
probability. Then one can check if this shape is consistent with
an even function of $L$, or if data require the presence of $L$-odd
terms. However, to map out the $L$ dependence for the disappearance
channel, one would need several data points, covering at least 1st and
2nd oscillation maxima. Therefore, currently such an analysis seems
not feasible with the proposed long-baseline experiments. The test
studied in Ref.~\cite{Schwetz:2021cuj} is based on the interplay of
disappearance and appearance channel, and therefore works already with
4 baselines (including the near detector). We leave for future
studies whether the disappearance test could potentially be performed
with atmospheric neutrinos.

\section{Estimation of non-constant density corrections}
\label{sec:corrections}

\begin{figure}
	\centering
	\includegraphics[width=0.4\textwidth]{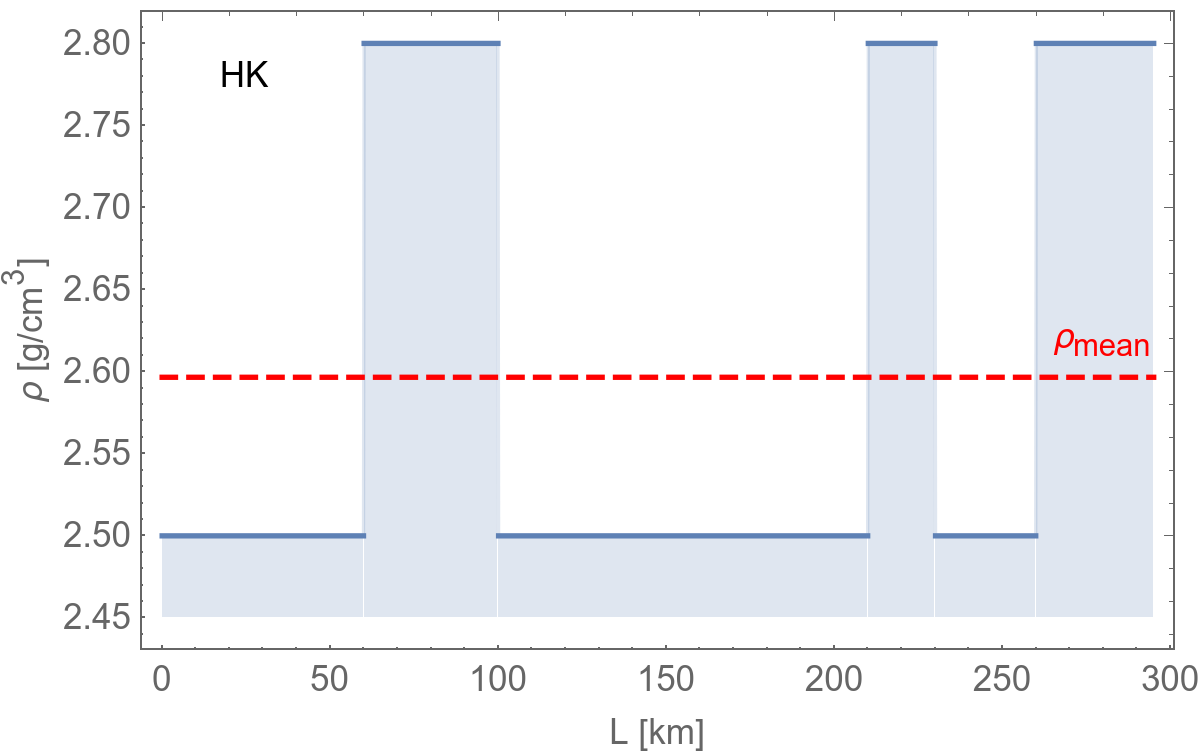} 
	\includegraphics[width=0.4\textwidth]{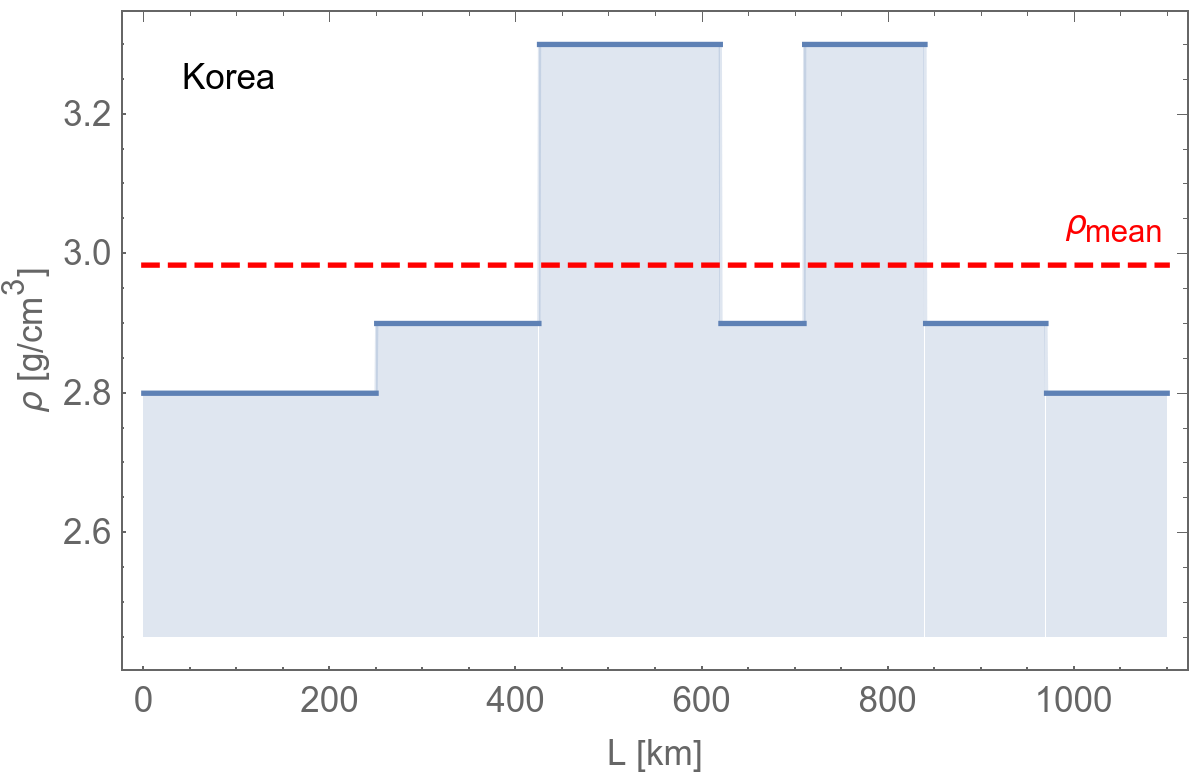} 
	\includegraphics[width=0.4\textwidth]{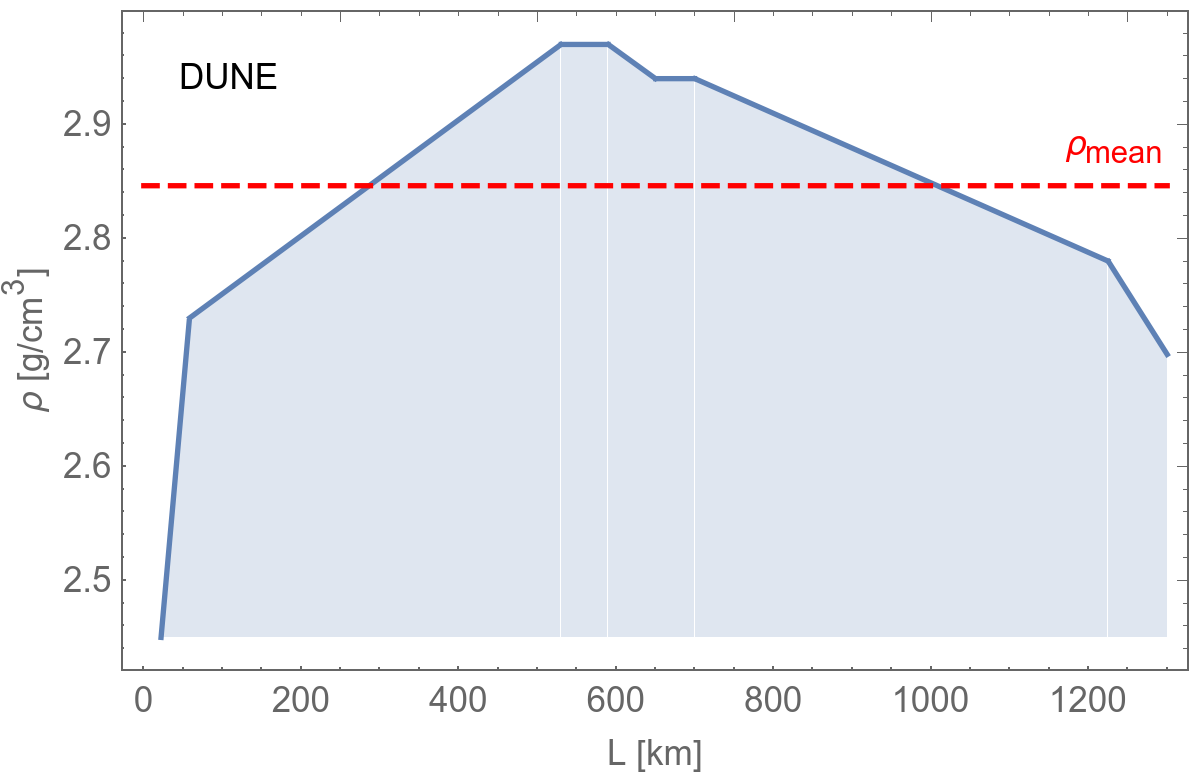}
	\caption{Matter density along the baseline of the T2HK (top)
          T2HKK (middle) and DUNE (bottom) experiments.
		Data taken from Refs.~\cite{Hagiwara:2011kw, Roe:2017zdw}.}
	\label{fig:density_profiles}
\end{figure}

In this section we are going to use this formalism to estimate the
impact of a non-constant density for our T violation test.
We will address the following two points:
\begin{enumerate}
\item when the average matter densities for different baselines are not exactly the same, and
\item an asymmetric density profile at a given baseline.  
\end{enumerate}
We consider these two cases using existing density profile studies for the 
T2HK(K)~\cite{Hagiwara:2011kw} and DUNE~\cite{Roe:2017zdw} experiments.\footnote{As shown in Ref.~\cite{Schwetz:2021cuj}, the ESS$\nu$SB experiment contributes only very little to the sensitivity of the T violation test. Therefore, we focus here on the DUNE and T2HK(K) experiments.}
The corresponding density profiles are shown in Fig.~\ref{fig:density_profiles}.

Since both non-constant density effects and new physics are small
corrections to the standard scenario with constant matter density, we
work in leading order in small effects: we can neglect any new
physics effect in estimating the size of the matter density corrections. In
this way, the non-constant density will introduce calculable
corrections, which can be taken into account in the T violation test,
as we describe below.

\subsection{Different average densities}
\label{sec:corr_average}

Let us first assume that the matter density can be considered constant
for each experiment, even if the average values are not the same.
From the profiles shown in
Fig.~\ref{fig:density_profiles}
we obtain
\begin{align}
  \bar\rho^{}_\mathrm{HK} &= 2.6~\mathrm{g/cm}^3 \,,\nonumber\\
  \bar\rho^{}_\mathrm{DUNE} &= 2.85~\mathrm{g/cm}^3 \,, \label{eq:mean_densities}\\  
  \bar\rho^{}_\mathrm{HKK} &= 3.0~\mathrm{g/cm}^3 \,.\nonumber
\end{align}
The Hamiltonian of the system is thus reduced to the time-independent $H_0$ in Eq.~\eqref{eq:H0}.
Assuming the standard neutrino model, it reads
\begin{equation}
	H_b = \frac{1}{2E}\, U
		\mqty[0 &&\\ &\Delta m^2_{21}&\\ &&\Delta m^2_{31}]
		U^\dagger
		+
		\mqty[ \bar v_b^{}&&\\ &0& \\ &&0]
\end{equation}
in the flavour basis, where $E$ is the neutrino energy, $U$ is the
standard PMNS mixing matrix, $\Delta m^2_{ij} \equiv m^2_i - m^2_j$
are the neutrino mass-squared differences, and
\begin{equation}
	v(\rho) = \sqrt{2} G_F n_e(\rho) \approx 
	3.78 \times 10^{-14}~\mathrm{eV}\, \left[\frac{\rho}{\mathrm{g/cm}^3}\right]\,.
\end{equation}
$\bar v_b$ denotes the potential corresponding to the average density
$\bar\rho_b$ and the index $b$ labels the different baselines, which
emphasizes that each baseline may have a different mean density, and
thus a different Hamiltonian.  We study the time evolution of the
system by diagonalizing this Hamiltonian numerically, which leads to a
set of effective masses $m^2_b = 2E\lambda_b$ and mixings $N^s_b =
N^d_b = W_b$ for each experiment.

\begin{figure}
	\centering
	\includegraphics[width=0.4\textwidth]{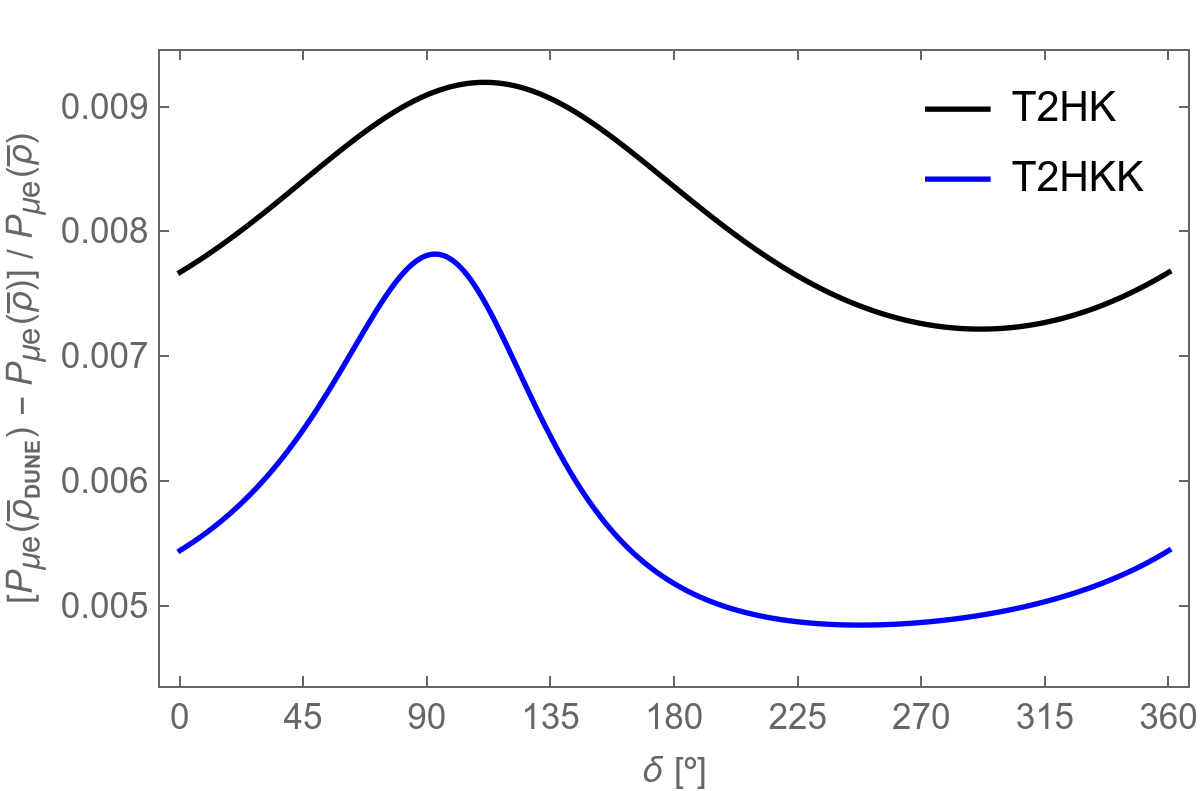}\\[2mm]
	\includegraphics[width=0.4\textwidth]{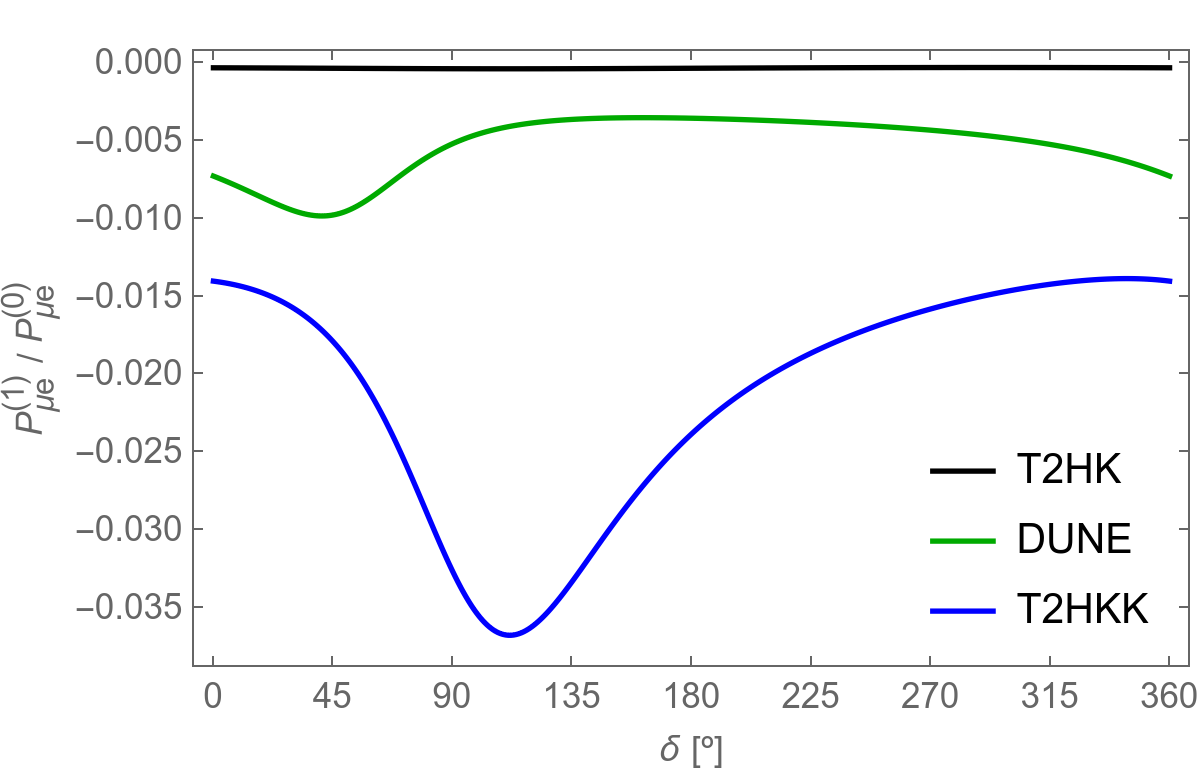} 
	\caption{Relative correction of the standard appearance
          probabilities as a function of the CP phase $\delta$ for $E=0.75$~GeV. The
          upper panel shows the relative error for T2HK and T2HKK if
          the constant mean density of 2.85~g/cm$^3$ is assumed
          instead of the correct ones according to
          Eq.~\eqref{eq:mean_densities}. The lower panel shows the
          relative size of the correction due to the non-constant
          matter potential $V(t)$.}
	\label{fig:corrections}
\end{figure}

In Tab.~\ref{tab:Psrhos} we give the values of disappearance and
appearance oscillation probabilities for the three baselines, assuming
different values for the mean densities for a few choices of the CP
phase $\delta$. The relative size of the effect for the appearance
probabilities is shown in the upper panel of
Fig.~\ref{fig:corrections}.  For the table and the figure we have
chosen the neutrino energy $E=0.75$~GeV, which has been found to
provide the best sensitivity in Ref.~\cite{Schwetz:2021cuj}.  For the
figure we assume the mean density $\bar\rho = 2.85~\mathrm{g/cm}^3$,
corresponding to the DUNE experiment, and show the relative error
induced for the T2HK and T2HKK baselines. As mentioned above, we can
use the standard oscillation scenario to estimate this effect,
considering only leading order terms in density variations and new
physics. We show the size of the correction as a function of the CP
phase $\delta$; all other oscillation parameters are set to their best
fit values~\cite{Esteban:2020cvm}. We see the effect is below 1\% for
all values of the CP phase $\delta$. For the disappearance
probabilities the effect is even smaller, compare
Tab.~\ref{tab:Psrhos}.

\begin{table}[t]
	\caption{Disappearance (left) and appearance (right) oscillation probabilities at the T2HK, T2HKK, and DUNE baselines, assuming different mean densities, for $\delta = 0,90^\circ,180^\circ$ and $E = 0.75$~GeV. The bold values correspond to the correct $\bar\rho$ for each experiment.}
	\label{tab:Psrhos}
	\begin{center}
		\begin{tabular}{rrr|ccc}
                  \hline\hline
			\multicolumn{2}{c}{$P_{\mu \mu}~(\%)$}&&\multicolumn{3}{c}{$\bar\rho~(\mathrm{g/cm}^3)$} \\ 
			\multicolumn{2}{c}{$L$~(km)} &&2.6	&2.85	&3.0			\\
			\hline
			\parbox[t]{3mm}{\multirow{3}{*}{\rotatebox[origin=c]{90}{$\delta=0$}}}
			&295	&&\textbf{11.69}&11.68			&11.68			\\	
			&1100	&&2.19			&2.19			&\textbf{2.19}	\\	
			&1300	&&41.77			&\textbf{41.78}	&41.78			\\
	                \hline                        
			\parbox[t]{3mm}{\multirow{3}{*}{\rotatebox[origin=c]{90}{$\delta=90^\circ$}}}
			&295	&&\textbf{12.05}&12.05			&12.05			\\	
			&1100	&&2.82			&2.82			&\textbf{2.82}	\\	
			&1300	&&38.94			&\textbf{38.91}	&38.89			\\
	                \hline
			\parbox[t]{3mm}{\multirow{3}{*}{\rotatebox[origin=c]{90}{$\delta=180^\circ$}}}
			&295	&&\textbf{12.41}&12.41			&12.41			\\	
			&1100	&&3.53			&3.53			&\textbf{3.53}	\\	
			&1300	&&36.22			&\textbf{36.15}	&36.12			\\
	                \hline\hline
		\end{tabular}
		~
		\begin{tabular}{rrr|ccc}
                  \hline\hline
			\multicolumn{2}{c}{$P_{\mu e}~(\%)$}&&\multicolumn{3}{c}{$\bar\rho~(\mathrm{g/cm}^3)$} \\ 
			\multicolumn{2}{c}{$L$~(km)}&
					&2.6			&2.85			&3.0			\\
			\hline
			\parbox[t]{3mm}{\multirow{3}{*}{\rotatebox[origin=c]{90}{$\delta=0$}}}
			&295	&&\textbf{4.77}	&4.80			&4.81			\\	
			&1100	&&5.77			&5.72			&\textbf{5.69}	\\	
			&1300	&&2.65			&\textbf{2.76}	&2.83			\\
	                \hline
			\parbox[t]{3mm}{\multirow{3}{*}{\rotatebox[origin=c]{90}{$\delta=90^\circ$}}}
			&295	&&\textbf{3.53}	&3.55			&3.56			\\	
			&1100	&&1.63			&1.62			&\textbf{1.60}	\\	
			&1300	&&2.06			&\textbf{2.17}	&2.23			\\
	                \hline 
			\parbox[t]{3mm}{\multirow{3}{*}{\rotatebox[origin=c]{90}{$\delta=180^\circ$}}}
			&295	&&\textbf{4.11}	&4.13			&4.15			\\	
			&1100	&&4.69			&4.65			&\textbf{4.63}	\\	
			&1300	&&7.93			&\textbf{8.12}	&8.24			\\
	                \hline\hline
		\end{tabular}
	\end{center}
\end{table}

Regarding the T violation test~\cite{Schwetz:2021cuj} based on
Eq.~\eqref{eq:ProbEven}, the main effect of the different mean
densities is that the fit parameters, the amplitudes $c_i$ and
frequencies $\omega_{ij}$, are no longer the same for all baselines.
Since the density dependence is small, we can include it in the fit
assuming
\begin{equation}
	c_i^b = \bar c_i + \delta c_i^b \,,
	\hspace{0.5cm}
	\omega_{ij}^b = \bar\omega_{ij} + \delta \omega_{ij}^b \,,
\end{equation}
with the reference parameters $\bar c_i$ and $\bar\omega_{ij}$
corresponding to standard oscillations and a common density $\bar\rho$
taken the same for all baselines.  Expanding up to first order in
these $L$-dependent perturbations, the $L$-even probability in
Eq.~\eqref{eq:ProbEven} becomes
\begin{align}
	P_{\alpha\beta}^{\mathrm{even},b} 
  	=& \sum_i \bar c_i^2 + 2\sum_{j<i} \bar c_i \bar c_j \cos(\bar \omega_{ij}L)
  	\nonumber\\
	 &+2\sum_{i} \bar c_i \delta c_i^b
	 +2\sum_{j<i}(\bar c_i \delta c_j^b + \bar c_j \delta c_i^b) \cos(\bar \omega_{ij}L)
  	\nonumber\\
	&-2\sum_{j<i} \bar c_i \bar c_j \delta \omega_{ij}^b L \sin(\bar \omega_{ij}L) \,.
        \label{eq:corr_mean}
\end{align}
Thus the crucial effect of different (mean) densities is the
appearance of new terms in the (previously) $L$-even oscillation
probability. Note the corrections shown in the 2nd and 3rd line of
Eq.~\eqref{eq:corr_mean} are known and fixed and can be just included
in the test described in Ref.~\cite{Schwetz:2021cuj} as constant
correction terms for each baseline. The fit itself can be performed
with an expression equivalent to the 1st line in
Eq.~\eqref{eq:corr_mean}, consistent with the leading order
perturbation approach mentioned above. From Fig.~\ref{fig:corrections}
we see that at $\delta = 0$ and $\pi$ relevant for the test, the
corrections are $\lesssim 0.5\%$ on the appearance probabilities, which
themselves are only few \%. Therefore, with realistic statistical
uncertainties these corrections are negligible.

\subsection{Non-constant density}

Let us now discuss the effect of a non-constant and non-symmetric
matter profile at a given baseline. We use the formalism developed
section~\ref{sec:T_nonconst} to calculate how much this affects the
probabilities to be probed in T2HK(K) and DUNE, assuming the standard
neutrino model (using the same perturbative argument as above).

Within our perturbation theory in $V(t)$, 
the zeroth-order result corresponds to the diagonalization of the Hamiltonian
with the constant mean density of the previous subsection.
Therefore,
the procedure described above yields 
the eigenvalues $\lambda_b$ and mixing matrices $N^s_b = N^d_b = W_b$ for each experiment.
As above,
we obtain the mean probabilities from Eq.~\eqref{eq:A0} as
$P_{\alpha\beta}^{(0)} = |\mathcal{A}_{\alpha\beta}^{(0)} |^2$.
The first-order correction to the oscillation amplitudes is then given by Eq.~\eqref{eq:A(1)}
in terms of these parameters and the matrix elements of the perturbation in the $H_0$ eigenbasis are
$V_{ij}(t) = W_{ei} W_{ej}^* \left[ v(t) - \bar v \right]$.
With this we can calculate the 1st order correction to the probabilities given in 
Eq.~\eqref{eq:Pcorr}. The relative size of this correction is shown in the lower panel of
Fig.~\ref{fig:corrections} as a function of $\delta$. We observe that
for T2HK these corrections are negligible and not visible on the scale
of the plot. For DUNE the effect is sub-percent for all values of
$\delta$. For T2HKK it can become as large as 3.5\% for $\delta \simeq
110^\circ$; for $\delta = 0$ and $\pi$ it is around 1.5\%. 
Similar as in sec.~\ref{sec:corr_average}, these are calculable and
fixed corrections to the probabilities, which can be taken into
account in the test proposed in Ref.~\cite{Schwetz:2021cuj}. However,
considering that these are percent-level correction on probabilities
which themselves are only a few \%, this effect is again negligibly
small, given realistic statistical errors.

\bigskip

Notice, however, that the size of such corrections does not directly
determine the amount of environmental T violation induced by the
matter profile. Even for the case of a symmetric profile with real
mixings, where no extra T violation is introduced, the oscillation
probabilities themselves get a non-vanishing correction. In order to
get a feeling for the size of environmental T violation at the DUNE
and T2HKK baselines we calculate the asymmetries~\eqref{eq:asym}
for the standard oscillation case considered above:
\begin{align}
  \text{T2HKK:} \quad &A_{\mu e}^{(1)} \approx  3.5\,(5.3)  \times 10^{-4} \,,\nonumber\\
  \text{DUNE:}  \quad &A_{\mu e}^{(1)} \approx -2.9\,(-2.0) \times 10^{-4} \,,
\end{align}
for $\delta = 0\,(180^\circ)$. These can be compared to the case of
$\delta = 90^\circ$, where we find 
\begin{align}
  \text{T2HKK:} \quad &A_{\mu e}^{(0)} \approx  7.1\, \times 10^{-2}
  \,, \quad 
  A_{\mu e}^{(1)} \approx -9.1\, \times 10^{-4} \,,\nonumber\\
  \text{DUNE:}  \quad &A_{\mu e}^{(0)} \approx  6.5\, \times 10^{-2}
  \,, \quad 
  A_{\mu e}^{(1)} \approx  3.8\, \times 10^{-4} \,.
\end{align}
We conclude that for these realistic
density profiles, environmental T violation is typically a \% level
effect compared to generic fundamental T violation~\cite{Hagiwara:2011kw, Roe:2017zdw}.

\section{Conclusions}
\label{sec:conclusion}

In this paper we have studied some aspects of the time reversal
transformation in a generic non-standard neutrino oscillation
framework. The motivation for our study is the model-independent T
violation test proposed recently in Ref.~\cite{Schwetz:2021cuj}. This
test can potentially be performed with three long-baseline
experiments, such as T2HK, DUNE and the proposed T2HKK. Here we
provide a theoretical discussion of the formalism for the
model-independent new physics parameterization proposed in
Ref.~\cite{Schwetz:2021cuj}. We derive the relevant flavour transition
amplitudes and probabilities and study their behaviour under the T
transformation.

The proposed parameterization covers a wide range of new-physics
scenarios in a model-independent way, including non-standard neutrino
interactions with arbitrary Lorentz structures in the charged and
neutral-current interaction, generic non-unitarity as well as sterile
neutrinos. We provide a discussion of fundamental versus
environmentally induced T violation, where the former is related to
complex phases in the theory while the latter is due to (standard or
non-standard) matter effects along the neutrino path. We show that a
result well known for standard oscillations holds also in our
extended scenario: in the absence of fundamental T violation,
environmental T violation can only be induced by an asymmetric matter
density profile.

We show that in general new-physics scenarios, also the disappearance
channel can be sensitive to T violating effects. This requires new
physics generating different flavour mixing at neutrino source and
detector. Although difficult to realise in practice, such an
observation offers in principle a clear signal of new physics, since
in the standard oscillation scenario no T violation is expected in the
disappearance channel.

Focusing on long-baseline accelerator neutrino experiments, we have
treated density variations along the neutrino path as a small
perturbation. Using detailed matter density profile studies for the
DUNE and T2HK(K) baselines from the literature, we have provided some
quantitative estimates on the corrections induced by a non-constant
matter density. Typically they are of order few percent or
smaller. Considering that appearance probabilities are themselves
typically only few percent, these corrections are much smaller than
realistic experimental uncertainties and hence, do not affect the test
proposed in Ref.~\cite{Schwetz:2021cuj}.

To conclude, the material presented here provides background
information to the model-independent T violation test from
Ref.~\cite{Schwetz:2021cuj}. This lies out the basis for the
possibility to test one of the fundamental symmetries of nature, the
time reversal symmetry, in a model-independent way using actually
planned neutrino oscillation experiments.

\bigskip\textbf{Acknowledgments.}
This project has received support from the European Union’s Horizon
2020 research and innovation programme under the Marie
Sklodowska-Curie grant agreement No 860881-HIDDeN,
and from the Alexander von Humboldt Foundation. 

\bibliography{Bibliography}

\begin{thebibliography}{54}%
\makeatletter
\providecommand \@ifxundefined [1]{%
 \@ifx{#1\undefined}
}%
\providecommand \@ifnum [1]{%
 \ifnum #1\expandafter \@firstoftwo
 \else \expandafter \@secondoftwo
 \fi
}%
\providecommand \@ifx [1]{%
 \ifx #1\expandafter \@firstoftwo
 \else \expandafter \@secondoftwo
 \fi
}%
\providecommand \natexlab [1]{#1}%
\providecommand \enquote  [1]{``#1''}%
\providecommand \bibnamefont  [1]{#1}%
\providecommand \bibfnamefont [1]{#1}%
\providecommand \citenamefont [1]{#1}%
\providecommand \href@noop [0]{\@secondoftwo}%
\providecommand \href [0]{\begingroup \@sanitize@url \@href}%
\providecommand \@href[1]{\@@startlink{#1}\@@href}%
\providecommand \@@href[1]{\endgroup#1\@@endlink}%
\providecommand \@sanitize@url [0]{\catcode `\\12\catcode `\$12\catcode
  `\&12\catcode `\#12\catcode `\^12\catcode `\_12\catcode `\%12\relax}%
\providecommand \@@startlink[1]{}%
\providecommand \@@endlink[0]{}%
\providecommand \url  [0]{\begingroup\@sanitize@url \@url }%
\providecommand \@url [1]{\endgroup\@href {#1}{\urlprefix }}%
\providecommand \urlprefix  [0]{URL }%
\providecommand \Eprint [0]{\href }%
\providecommand \doibase [0]{https://doi.org/}%
\providecommand \selectlanguage [0]{\@gobble}%
\providecommand \bibinfo  [0]{\@secondoftwo}%
\providecommand \bibfield  [0]{\@secondoftwo}%
\providecommand \translation [1]{[#1]}%
\providecommand \BibitemOpen [0]{}%
\providecommand \bibitemStop [0]{}%
\providecommand \bibitemNoStop [0]{.\EOS\space}%
\providecommand \EOS [0]{\spacefactor3000\relax}%
\providecommand \BibitemShut  [1]{\csname bibitem#1\endcsname}%
\let\auto@bib@innerbib\@empty
\bibitem [{\citenamefont {Schwetz}\ and\ \citenamefont
  {Segarra}(2021)}]{Schwetz:2021cuj}%
  \BibitemOpen
  \bibfield  {author} {\bibinfo {author} {\bibfnamefont {T.}~\bibnamefont
  {Schwetz}}\ and\ \bibinfo {author} {\bibfnamefont {A.}~\bibnamefont
  {Segarra}},\ }\bibfield  {title} {\bibinfo {title} {{Model-Independent Test
  of T Violation in Neutrino Oscillations}},\ }\href@noop {} {\  (\bibinfo
  {year} {2021})},\ \Eprint {https://arxiv.org/abs/2106.16099}
  {arXiv:2106.16099 [hep-ph]} \BibitemShut {NoStop}%
\bibitem [{\citenamefont {Cabibbo}(1978)}]{Cabibbo:1977nk}%
  \BibitemOpen
  \bibfield  {author} {\bibinfo {author} {\bibfnamefont {N.}~\bibnamefont
  {Cabibbo}},\ }\bibfield  {title} {\bibinfo {title} {{Time Reversal Violation
  in Neutrino Oscillation}},\ }\href
  {https://doi.org/10.1016/0370-2693(78)90132-6} {\bibfield  {journal}
  {\bibinfo  {journal} {Phys. Lett.}\ }\textbf {\bibinfo {volume} {72B}},\
  \bibinfo {pages} {333} (\bibinfo {year} {1978})}\BibitemShut {NoStop}%
\bibitem [{\citenamefont {Bilenky}\ \emph {et~al.}(1980)\citenamefont
  {Bilenky}, \citenamefont {Hosek},\ and\ \citenamefont
  {Petcov}}]{Bilenky:1980cx}%
  \BibitemOpen
  \bibfield  {author} {\bibinfo {author} {\bibfnamefont {S.~M.}\ \bibnamefont
  {Bilenky}}, \bibinfo {author} {\bibfnamefont {J.}~\bibnamefont {Hosek}},\
  and\ \bibinfo {author} {\bibfnamefont {S.~T.}\ \bibnamefont {Petcov}},\
  }\bibfield  {title} {\bibinfo {title} {{On Oscillations of Neutrinos with
  Dirac and Majorana Masses}},\ }\href
  {https://doi.org/10.1016/0370-2693(80)90927-2} {\bibfield  {journal}
  {\bibinfo  {journal} {Phys. Lett.}\ }\textbf {\bibinfo {volume} {94B}},\
  \bibinfo {pages} {495} (\bibinfo {year} {1980})}\BibitemShut {NoStop}%
\bibitem [{\citenamefont {Barger}\ \emph {et~al.}(1980)\citenamefont {Barger},
  \citenamefont {Whisnant},\ and\ \citenamefont {Phillips}}]{Barger:1980jm}%
  \BibitemOpen
  \bibfield  {author} {\bibinfo {author} {\bibfnamefont {V.~D.}\ \bibnamefont
  {Barger}}, \bibinfo {author} {\bibfnamefont {K.}~\bibnamefont {Whisnant}},\
  and\ \bibinfo {author} {\bibfnamefont {R.~J.~N.}\ \bibnamefont {Phillips}},\
  }\bibfield  {title} {\bibinfo {title} {{CP Violation in Three Neutrino
  Oscillations}},\ }\href {https://doi.org/10.1103/PhysRevLett.45.2084}
  {\bibfield  {journal} {\bibinfo  {journal} {Phys. Rev. Lett.}\ }\textbf
  {\bibinfo {volume} {45}},\ \bibinfo {pages} {2084} (\bibinfo {year}
  {1980})}\BibitemShut {NoStop}%
\bibitem [{\citenamefont {Zyla}\ \emph {et~al.}(2020)\citenamefont {Zyla} \emph
  {et~al.}}]{Zyla:2020zbs}%
  \BibitemOpen
  \bibfield  {author} {\bibinfo {author} {\bibfnamefont {P.}~\bibnamefont
  {Zyla}} \emph {et~al.} (\bibinfo {collaboration} {Particle Data Group}),\
  }\bibfield  {title} {\bibinfo {title} {{Review of Particle Physics}},\ }\href
  {https://doi.org/10.1093/ptep/ptaa104} {\bibfield  {journal} {\bibinfo
  {journal} {PTEP}\ }\textbf {\bibinfo {volume} {2020}},\ \bibinfo {pages}
  {083C01} (\bibinfo {year} {2020})}\BibitemShut {NoStop}%
\bibitem [{\citenamefont {Pontecorvo}(1968)}]{Pontecorvo:1967fh}%
  \BibitemOpen
  \bibfield  {author} {\bibinfo {author} {\bibfnamefont {B.}~\bibnamefont
  {Pontecorvo}},\ }\bibfield  {title} {\bibinfo {title} {{Neutrino Experiments
  and the Problem of Conservation of Leptonic Charge}},\ }\href@noop {}
  {\bibfield  {journal} {\bibinfo  {journal} {Sov. Phys. JETP}\ }\textbf
  {\bibinfo {volume} {26}},\ \bibinfo {pages} {984} (\bibinfo {year} {1968})},\
  \bibinfo {note} {[Zh. Eksp. Teor. Fiz.53,1717(1967)]}\BibitemShut {NoStop}%
\bibitem [{\citenamefont {Gribov}\ and\ \citenamefont
  {Pontecorvo}(1969)}]{Gribov:1968kq}%
  \BibitemOpen
  \bibfield  {author} {\bibinfo {author} {\bibfnamefont {V.~N.}\ \bibnamefont
  {Gribov}}\ and\ \bibinfo {author} {\bibfnamefont {B.}~\bibnamefont
  {Pontecorvo}},\ }\bibfield  {title} {\bibinfo {title} {{Neutrino Astronomy
  and Lepton Charge}},\ }\href {https://doi.org/10.1016/0370-2693(69)90525-5}
  {\bibfield  {journal} {\bibinfo  {journal} {Phys. Lett.}\ }\textbf {\bibinfo
  {volume} {B28}},\ \bibinfo {pages} {493} (\bibinfo {year}
  {1969})}\BibitemShut {NoStop}%
\bibitem [{\citenamefont {Maki}\ \emph {et~al.}(1962)\citenamefont {Maki},
  \citenamefont {Nakagawa},\ and\ \citenamefont {Sakata}}]{Maki:1962mu}%
  \BibitemOpen
  \bibfield  {author} {\bibinfo {author} {\bibfnamefont {Z.}~\bibnamefont
  {Maki}}, \bibinfo {author} {\bibfnamefont {M.}~\bibnamefont {Nakagawa}},\
  and\ \bibinfo {author} {\bibfnamefont {S.}~\bibnamefont {Sakata}},\
  }\bibfield  {title} {\bibinfo {title} {{Remarks on the Unified Model of
  Elementary Particles}},\ }\href {https://doi.org/10.1143/PTP.28.870}
  {\bibfield  {journal} {\bibinfo  {journal} {Prog. Theor. Phys.}\ }\textbf
  {\bibinfo {volume} {28}},\ \bibinfo {pages} {870} (\bibinfo {year}
  {1962})}\BibitemShut {NoStop}%
\bibitem [{\citenamefont {Kobayashi}\ and\ \citenamefont
  {Maskawa}(1973)}]{Kobayashi:1973fv}%
  \BibitemOpen
  \bibfield  {author} {\bibinfo {author} {\bibfnamefont {M.}~\bibnamefont
  {Kobayashi}}\ and\ \bibinfo {author} {\bibfnamefont {T.}~\bibnamefont
  {Maskawa}},\ }\bibfield  {title} {\bibinfo {title} {{CP Violation in the
  Renormalizable Theory of Weak Interaction}},\ }\href
  {https://doi.org/10.1143/PTP.49.652} {\bibfield  {journal} {\bibinfo
  {journal} {Prog. Theor. Phys.}\ }\textbf {\bibinfo {volume} {49}},\ \bibinfo
  {pages} {652} (\bibinfo {year} {1973})}\BibitemShut {NoStop}%
\bibitem [{\citenamefont {Wolfenstein}(1978)}]{Wolfenstein:1977ue}%
  \BibitemOpen
  \bibfield  {author} {\bibinfo {author} {\bibfnamefont {L.}~\bibnamefont
  {Wolfenstein}},\ }\bibfield  {title} {\bibinfo {title} {{Neutrino
  Oscillations in Matter}},\ }\href {https://doi.org/10.1103/PhysRevD.17.2369}
  {\bibfield  {journal} {\bibinfo  {journal} {Phys. Rev.}\ }\textbf {\bibinfo
  {volume} {D17}},\ \bibinfo {pages} {2369} (\bibinfo {year}
  {1978})}\BibitemShut {NoStop}%
\bibitem [{\citenamefont {Langacker}\ \emph {et~al.}(1987)\citenamefont
  {Langacker}, \citenamefont {Petcov}, \citenamefont {Steigman},\ and\
  \citenamefont {Toshev}}]{Langacker:1986jv}%
  \BibitemOpen
  \bibfield  {author} {\bibinfo {author} {\bibfnamefont {P.}~\bibnamefont
  {Langacker}}, \bibinfo {author} {\bibfnamefont {S.~T.}\ \bibnamefont
  {Petcov}}, \bibinfo {author} {\bibfnamefont {G.}~\bibnamefont {Steigman}},\
  and\ \bibinfo {author} {\bibfnamefont {S.}~\bibnamefont {Toshev}},\
  }\bibfield  {title} {\bibinfo {title} {{On the Mikheev-Smirnov-Wolfenstein
  (MSW) Mechanism of Amplification of Neutrino Oscillations in Matter}},\
  }\href {https://doi.org/10.1016/0550-3213(87)90699-7} {\bibfield  {journal}
  {\bibinfo  {journal} {Nucl. Phys.}\ }\textbf {\bibinfo {volume} {B282}},\
  \bibinfo {pages} {589} (\bibinfo {year} {1987})}\BibitemShut {NoStop}%
\bibitem [{\citenamefont {Bernab\'eu}\ and\ \citenamefont
  {Segarra}(2018)}]{Bernabeu:2018use}%
  \BibitemOpen
  \bibfield  {author} {\bibinfo {author} {\bibfnamefont {J.}~\bibnamefont
  {Bernab\'eu}}\ and\ \bibinfo {author} {\bibfnamefont {A.}~\bibnamefont
  {Segarra}},\ }\bibfield  {title} {\bibinfo {title} {{Signatures of the
  genuine and matter-induced components of the CP violation asymmetry in
  neutrino oscillations}},\ }\href {https://doi.org/10.1007/JHEP11(2018)063}
  {\bibfield  {journal} {\bibinfo  {journal} {JHEP}\ }\textbf {\bibinfo
  {volume} {11}},\ \bibinfo {pages} {063}},\ \Eprint
  {https://arxiv.org/abs/1807.11879} {arXiv:1807.11879 [hep-ph]} \BibitemShut
  {NoStop}%
\bibitem [{\citenamefont {Bernab\'eu}\ and\ \citenamefont
  {Segarra}(2019)}]{Bernabeu:2019npc}%
  \BibitemOpen
  \bibfield  {author} {\bibinfo {author} {\bibfnamefont {J.}~\bibnamefont
  {Bernab\'eu}}\ and\ \bibinfo {author} {\bibfnamefont {A.}~\bibnamefont
  {Segarra}},\ }\bibfield  {title} {\bibinfo {title} {{Do T asymmetries for
  neutrino oscillations in uniform matter have a CP-even component?}},\ }\href
  {https://doi.org/10.1007/JHEP03(2019)103} {\bibfield  {journal} {\bibinfo
  {journal} {JHEP}\ }\textbf {\bibinfo {volume} {03}},\ \bibinfo {pages}
  {103}},\ \Eprint {https://arxiv.org/abs/1901.02761} {arXiv:1901.02761
  [hep-ph]} \BibitemShut {NoStop}%
\bibitem [{\citenamefont {Krastev}\ and\ \citenamefont
  {Petcov}(1988)}]{Krastev:1988yu}%
  \BibitemOpen
  \bibfield  {author} {\bibinfo {author} {\bibfnamefont {P.}~\bibnamefont
  {Krastev}}\ and\ \bibinfo {author} {\bibfnamefont {S.}~\bibnamefont
  {Petcov}},\ }\bibfield  {title} {\bibinfo {title} {{Resonance Amplification
  and T Violation Effects in Three Neutrino Oscillations in the Earth}},\
  }\href {https://doi.org/10.1016/0370-2693(88)90404-2} {\bibfield  {journal}
  {\bibinfo  {journal} {Phys. Lett. B}\ }\textbf {\bibinfo {volume} {205}},\
  \bibinfo {pages} {84} (\bibinfo {year} {1988})}\BibitemShut {NoStop}%
\bibitem [{\citenamefont {Akhmedov}\ \emph {et~al.}(2001)\citenamefont
  {Akhmedov}, \citenamefont {Huber}, \citenamefont {Lindner},\ and\
  \citenamefont {Ohlsson}}]{Akhmedov:2001kd}%
  \BibitemOpen
  \bibfield  {author} {\bibinfo {author} {\bibfnamefont {E.~K.}\ \bibnamefont
  {Akhmedov}}, \bibinfo {author} {\bibfnamefont {P.}~\bibnamefont {Huber}},
  \bibinfo {author} {\bibfnamefont {M.}~\bibnamefont {Lindner}},\ and\ \bibinfo
  {author} {\bibfnamefont {T.}~\bibnamefont {Ohlsson}},\ }\bibfield  {title}
  {\bibinfo {title} {{T Violation in Neutrino Oscillations in Matter}},\ }\href
  {https://doi.org/10.1016/S0550-3213(01)00261-9} {\bibfield  {journal}
  {\bibinfo  {journal} {Nucl. Phys. B}\ }\textbf {\bibinfo {volume} {608}},\
  \bibinfo {pages} {394} (\bibinfo {year} {2001})},\ \Eprint
  {https://arxiv.org/abs/hep-ph/0105029} {arXiv:hep-ph/0105029} \BibitemShut
  {NoStop}%
\bibitem [{\citenamefont {Kuo}\ and\ \citenamefont
  {Pantaleone}(1987)}]{Kuo:1987km}%
  \BibitemOpen
  \bibfield  {author} {\bibinfo {author} {\bibfnamefont {T.-K.}\ \bibnamefont
  {Kuo}}\ and\ \bibinfo {author} {\bibfnamefont {J.~T.}\ \bibnamefont
  {Pantaleone}},\ }\bibfield  {title} {\bibinfo {title} {{$T$ Nonconservation
  in Three Neutrino Oscillations}},\ }\href
  {https://doi.org/10.1016/0370-2693(87)90688-5} {\bibfield  {journal}
  {\bibinfo  {journal} {Phys. Lett. B}\ }\textbf {\bibinfo {volume} {198}},\
  \bibinfo {pages} {406} (\bibinfo {year} {1987})}\BibitemShut {NoStop}%
\bibitem [{\citenamefont {Toshev}(1989)}]{Toshev:1989vz}%
  \BibitemOpen
  \bibfield  {author} {\bibinfo {author} {\bibfnamefont {S.}~\bibnamefont
  {Toshev}},\ }\bibfield  {title} {\bibinfo {title} {{Maximal $T$ Violation in
  Matter}},\ }\href {https://doi.org/10.1016/0370-2693(89)91205-7} {\bibfield
  {journal} {\bibinfo  {journal} {Phys. Lett. B}\ }\textbf {\bibinfo {volume}
  {226}},\ \bibinfo {pages} {335} (\bibinfo {year} {1989})}\BibitemShut
  {NoStop}%
\bibitem [{\citenamefont {Toshev}(1991)}]{Toshev:1991ku}%
  \BibitemOpen
  \bibfield  {author} {\bibinfo {author} {\bibfnamefont {S.}~\bibnamefont
  {Toshev}},\ }\bibfield  {title} {\bibinfo {title} {{On T Violation in Matter
  Neutrino Oscillations}},\ }\href {https://doi.org/10.1142/S0217732391000464}
  {\bibfield  {journal} {\bibinfo  {journal} {Mod. Phys. Lett. A}\ }\textbf
  {\bibinfo {volume} {6}},\ \bibinfo {pages} {455} (\bibinfo {year}
  {1991})}\BibitemShut {NoStop}%
\bibitem [{\citenamefont {Arafune}\ and\ \citenamefont
  {Sato}(1997)}]{Arafune:1996bt}%
  \BibitemOpen
  \bibfield  {author} {\bibinfo {author} {\bibfnamefont {J.}~\bibnamefont
  {Arafune}}\ and\ \bibinfo {author} {\bibfnamefont {J.}~\bibnamefont {Sato}},\
  }\bibfield  {title} {\bibinfo {title} {{CP and T Violation Test in Neutrino
  Oscillation}},\ }\href {https://doi.org/10.1103/PhysRevD.55.1653} {\bibfield
  {journal} {\bibinfo  {journal} {Phys. Rev. D}\ }\textbf {\bibinfo {volume}
  {55}},\ \bibinfo {pages} {1653} (\bibinfo {year} {1997})},\ \Eprint
  {https://arxiv.org/abs/hep-ph/9607437} {arXiv:hep-ph/9607437} \BibitemShut
  {NoStop}%
\bibitem [{\citenamefont {Parke}\ and\ \citenamefont
  {Weiler}(2001)}]{Parke:2000hu}%
  \BibitemOpen
  \bibfield  {author} {\bibinfo {author} {\bibfnamefont {S.~J.}\ \bibnamefont
  {Parke}}\ and\ \bibinfo {author} {\bibfnamefont {T.~J.}\ \bibnamefont
  {Weiler}},\ }\bibfield  {title} {\bibinfo {title} {{Optimizing T Violating
  Effects for Neutrino Oscillations in Matter}},\ }\href
  {https://doi.org/10.1016/S0370-2693(01)00111-3} {\bibfield  {journal}
  {\bibinfo  {journal} {Phys. Lett. B}\ }\textbf {\bibinfo {volume} {501}},\
  \bibinfo {pages} {106} (\bibinfo {year} {2001})},\ \Eprint
  {https://arxiv.org/abs/hep-ph/0011247} {arXiv:hep-ph/0011247} \BibitemShut
  {NoStop}%
\bibitem [{\citenamefont {Xing}(2013)}]{Xing:2013uxa}%
  \BibitemOpen
  \bibfield  {author} {\bibinfo {author} {\bibfnamefont {Z.-z.}\ \bibnamefont
  {Xing}},\ }\bibfield  {title} {\bibinfo {title} {{Leptonic Commutators and
  Clean T Violation in Neutrino Oscillations}},\ }\href
  {https://doi.org/10.1103/PhysRevD.88.017301} {\bibfield  {journal} {\bibinfo
  {journal} {Phys. Rev. D}\ }\textbf {\bibinfo {volume} {88}},\ \bibinfo
  {pages} {017301} (\bibinfo {year} {2013})},\ \Eprint
  {https://arxiv.org/abs/1304.7606} {arXiv:1304.7606 [hep-ph]} \BibitemShut
  {NoStop}%
\bibitem [{\citenamefont {Petcov}\ and\ \citenamefont
  {Zhou}(2018)}]{Petcov:2018zka}%
  \BibitemOpen
  \bibfield  {author} {\bibinfo {author} {\bibfnamefont {S.~T.}\ \bibnamefont
  {Petcov}}\ and\ \bibinfo {author} {\bibfnamefont {Y.-L.}\ \bibnamefont
  {Zhou}},\ }\bibfield  {title} {\bibinfo {title} {{On Neutrino Mixing in
  Matter and CP and T Violation Effects in Neutrino Oscillations}},\ }\href
  {https://doi.org/10.1016/j.physletb.2018.08.025} {\bibfield  {journal}
  {\bibinfo  {journal} {Phys. Lett. B}\ }\textbf {\bibinfo {volume} {785}},\
  \bibinfo {pages} {95} (\bibinfo {year} {2018})},\ \Eprint
  {https://arxiv.org/abs/1806.09112} {arXiv:1806.09112 [hep-ph]} \BibitemShut
  {NoStop}%
\bibitem [{\citenamefont {Abe}\ \emph {et~al.}(2020)\citenamefont {Abe} \emph
  {et~al.}}]{Abe:2019vii}%
  \BibitemOpen
  \bibfield  {author} {\bibinfo {author} {\bibfnamefont {K.}~\bibnamefont
  {Abe}} \emph {et~al.} (\bibinfo {collaboration} {T2K}),\ }\bibfield  {title}
  {\bibinfo {title} {{Constraint on the Matter--Antimatter Symmetry-Violating
  Phase in Neutrino Oscillations}},\ }\href
  {https://doi.org/10.1038/s41586-020-2177-0} {\bibfield  {journal} {\bibinfo
  {journal} {Nature}\ }\textbf {\bibinfo {volume} {580}},\ \bibinfo {pages}
  {339} (\bibinfo {year} {2020})},\ \bibinfo {note} {[Erratum: Nature 583,E
  \textbf{16} (2020)]},\ \Eprint {https://arxiv.org/abs/1910.03887}
  {arXiv:1910.03887 [hep-ex]} \BibitemShut {NoStop}%
\bibitem [{\citenamefont {Acero}\ \emph {et~al.}(2019)\citenamefont {Acero}
  \emph {et~al.}}]{Acero:2019ksn}%
  \BibitemOpen
  \bibfield  {author} {\bibinfo {author} {\bibfnamefont {M.}~\bibnamefont
  {Acero}} \emph {et~al.} (\bibinfo {collaboration} {NOvA}),\ }\bibfield
  {title} {\bibinfo {title} {{First Measurement of Neutrino Oscillation
  Parameters Using Neutrinos and Antineutrinos by Nova}},\ }\href
  {https://doi.org/10.1103/PhysRevLett.123.151803} {\bibfield  {journal}
  {\bibinfo  {journal} {Phys. Rev. Lett.}\ }\textbf {\bibinfo {volume} {123}},\
  \bibinfo {pages} {151803} (\bibinfo {year} {2019})},\ \Eprint
  {https://arxiv.org/abs/1906.04907} {arXiv:1906.04907 [hep-ex]} \BibitemShut
  {NoStop}%
\bibitem [{\citenamefont {Esteban}\ \emph {et~al.}(2020)\citenamefont
  {Esteban}, \citenamefont {Gonzalez-Garcia}, \citenamefont {Maltoni},
  \citenamefont {Schwetz},\ and\ \citenamefont {Zhou}}]{Esteban:2020cvm}%
  \BibitemOpen
  \bibfield  {author} {\bibinfo {author} {\bibfnamefont {I.}~\bibnamefont
  {Esteban}}, \bibinfo {author} {\bibfnamefont {M.~C.}\ \bibnamefont
  {Gonzalez-Garcia}}, \bibinfo {author} {\bibfnamefont {M.}~\bibnamefont
  {Maltoni}}, \bibinfo {author} {\bibfnamefont {T.}~\bibnamefont {Schwetz}},\
  and\ \bibinfo {author} {\bibfnamefont {A.}~\bibnamefont {Zhou}},\ }\bibfield
  {title} {\bibinfo {title} {{The fate of hints: updated global analysis of
  three-flavor neutrino oscillations}},\ }\href
  {https://doi.org/10.1007/JHEP09(2020)178} {\bibfield  {journal} {\bibinfo
  {journal} {JHEP}\ }\textbf {\bibinfo {volume} {09}},\ \bibinfo {pages}
  {178}},\ \Eprint {https://arxiv.org/abs/2007.14792} {arXiv:2007.14792
  [hep-ph]} \BibitemShut {NoStop}%
\bibitem [{\citenamefont {de~Salas}\ \emph {et~al.}(2020)\citenamefont
  {de~Salas}, \citenamefont {Forero}, \citenamefont {Gariazzo}, \citenamefont
  {Martínez-Miravé}, \citenamefont {Mena}, \citenamefont {Ternes},
  \citenamefont {Tórtola},\ and\ \citenamefont {Valle}}]{deSalas:2020pgw}%
  \BibitemOpen
  \bibfield  {author} {\bibinfo {author} {\bibfnamefont {P.}~\bibnamefont
  {de~Salas}}, \bibinfo {author} {\bibfnamefont {D.}~\bibnamefont {Forero}},
  \bibinfo {author} {\bibfnamefont {S.}~\bibnamefont {Gariazzo}}, \bibinfo
  {author} {\bibfnamefont {P.}~\bibnamefont {Martínez-Miravé}}, \bibinfo
  {author} {\bibfnamefont {O.}~\bibnamefont {Mena}}, \bibinfo {author}
  {\bibfnamefont {C.}~\bibnamefont {Ternes}}, \bibinfo {author} {\bibfnamefont
  {M.}~\bibnamefont {Tórtola}},\ and\ \bibinfo {author} {\bibfnamefont
  {J.}~\bibnamefont {Valle}},\ }\bibfield  {title} {\bibinfo {title} {{2020
  Global Reassessment of the Neutrino Oscillation Picture}},\ }\href@noop {} {\
   (\bibinfo {year} {2020})},\ \Eprint {https://arxiv.org/abs/2006.11237}
  {arXiv:2006.11237 [hep-ph]} \BibitemShut {NoStop}%
\bibitem [{\citenamefont {Capozzi}\ \emph {et~al.}(2021)\citenamefont
  {Capozzi}, \citenamefont {Di~Valentino}, \citenamefont {Lisi}, \citenamefont
  {Marrone}, \citenamefont {Melchiorri},\ and\ \citenamefont
  {Palazzo}}]{Capozzi:2021fjo}%
  \BibitemOpen
  \bibfield  {author} {\bibinfo {author} {\bibfnamefont {F.}~\bibnamefont
  {Capozzi}}, \bibinfo {author} {\bibfnamefont {E.}~\bibnamefont
  {Di~Valentino}}, \bibinfo {author} {\bibfnamefont {E.}~\bibnamefont {Lisi}},
  \bibinfo {author} {\bibfnamefont {A.}~\bibnamefont {Marrone}}, \bibinfo
  {author} {\bibfnamefont {A.}~\bibnamefont {Melchiorri}},\ and\ \bibinfo
  {author} {\bibfnamefont {A.}~\bibnamefont {Palazzo}},\ }\bibfield  {title}
  {\bibinfo {title} {{Unfinished Fabric of the Three Neutrino Paradigm}},\
  }\href {https://doi.org/10.1103/PhysRevD.104.083031} {\bibfield  {journal}
  {\bibinfo  {journal} {Phys. Rev. D}\ }\textbf {\bibinfo {volume} {104}},\
  \bibinfo {pages} {083031} (\bibinfo {year} {2021})},\ \Eprint
  {https://arxiv.org/abs/2107.00532} {arXiv:2107.00532 [hep-ph]} \BibitemShut
  {NoStop}%
\bibitem [{\citenamefont {Abi}\ \emph {et~al.}(2020{\natexlab{a}})\citenamefont
  {Abi} \emph {et~al.}}]{Abi:2020wmh}%
  \BibitemOpen
  \bibfield  {author} {\bibinfo {author} {\bibfnamefont {B.}~\bibnamefont
  {Abi}} \emph {et~al.} (\bibinfo {collaboration} {DUNE}),\ }\bibfield  {title}
  {\bibinfo {title} {{Deep Underground Neutrino Experiment (Dune), Far Detector
  Technical Design Report, Volume I Introduction to Dune}},\ }\href
  {https://doi.org/10.1088/1748-0221/15/08/T08008} {\bibfield  {journal}
  {\bibinfo  {journal} {JINST}\ }\textbf {\bibinfo {volume} {15}}\bibfield
  {number} {\bibinfo  {number} { (08)},\ \bibinfo {pages} {T08008}},\ }\Eprint
  {https://arxiv.org/abs/2002.02967} {arXiv:2002.02967 [physics.ins-det]}
  \BibitemShut {NoStop}%
\bibitem [{\citenamefont {Abi}\ \emph {et~al.}(2020{\natexlab{b}})\citenamefont
  {Abi} \emph {et~al.}}]{Abi:2020evt}%
  \BibitemOpen
  \bibfield  {author} {\bibinfo {author} {\bibfnamefont {B.}~\bibnamefont
  {Abi}} \emph {et~al.} (\bibinfo {collaboration} {DUNE}),\ }\bibfield  {title}
  {\bibinfo {title} {{Deep Underground Neutrino Experiment (Dune), Far Detector
  Technical Design Report, Volume II: Dune Physics}},\ }\href@noop {} {\
  (\bibinfo {year} {2020}{\natexlab{b}})},\ \Eprint
  {https://arxiv.org/abs/2002.03005} {arXiv:2002.03005 [hep-ex]} \BibitemShut
  {NoStop}%
\bibitem [{\citenamefont {Abe}\ \emph {et~al.}(2018{\natexlab{a}})\citenamefont
  {Abe} \emph {et~al.}}]{Abe:2018uyc}%
  \BibitemOpen
  \bibfield  {author} {\bibinfo {author} {\bibfnamefont {K.}~\bibnamefont
  {Abe}} \emph {et~al.} (\bibinfo {collaboration} {Hyper-Kamiokande}),\
  }\href@noop {} {\bibinfo {title} {{Hyper-Kamiokande Design Report}}}
  (\bibinfo {year} {2018}{\natexlab{a}}),\ \Eprint
  {https://arxiv.org/abs/1805.04163} {arXiv:1805.04163 [physics.ins-det]}
  \BibitemShut {NoStop}%
\bibitem [{\citenamefont {Abe}\ \emph {et~al.}(2018{\natexlab{b}})\citenamefont
  {Abe} \emph {et~al.}}]{Abe:2016ero}%
  \BibitemOpen
  \bibfield  {author} {\bibinfo {author} {\bibfnamefont {K.}~\bibnamefont
  {Abe}} \emph {et~al.} (\bibinfo {collaboration} {Hyper-Kamiokande}),\
  }\bibfield  {title} {\bibinfo {title} {{Physics potentials with the second
  Hyper-Kamiokande detector in Korea}},\ }\href
  {https://doi.org/10.1093/ptep/pty044} {\bibfield  {journal} {\bibinfo
  {journal} {PTEP}\ }\textbf {\bibinfo {volume} {2018}},\ \bibinfo {pages}
  {063C01} (\bibinfo {year} {2018}{\natexlab{b}})},\ \Eprint
  {https://arxiv.org/abs/1611.06118} {arXiv:1611.06118 [hep-ex]} \BibitemShut
  {NoStop}%
\bibitem [{\citenamefont {Baussan}\ \emph {et~al.}(2014)\citenamefont {Baussan}
  \emph {et~al.}}]{Baussan:2013zcy}%
  \BibitemOpen
  \bibfield  {author} {\bibinfo {author} {\bibfnamefont {E.}~\bibnamefont
  {Baussan}} \emph {et~al.} (\bibinfo {collaboration} {ESSnuSB}),\ }\bibfield
  {title} {\bibinfo {title} {{A Very Intense Neutrino Super Beam Experiment for
  Leptonic CP Violation Discovery Based on the European Spallation Source
  Linac}},\ }\href {https://doi.org/10.1016/j.nuclphysb.2014.05.016} {\bibfield
   {journal} {\bibinfo  {journal} {Nucl. Phys. B}\ }\textbf {\bibinfo {volume}
  {885}},\ \bibinfo {pages} {127} (\bibinfo {year} {2014})},\ \Eprint
  {https://arxiv.org/abs/1309.7022} {arXiv:1309.7022 [hep-ex]} \BibitemShut
  {NoStop}%
\bibitem [{\citenamefont {Blennow}\ \emph {et~al.}(2020)\citenamefont
  {Blennow}, \citenamefont {Fernandez-Martinez}, \citenamefont {Ota},\ and\
  \citenamefont {Rosauro-Alcaraz}}]{Blennow:2019bvl}%
  \BibitemOpen
  \bibfield  {author} {\bibinfo {author} {\bibfnamefont {M.}~\bibnamefont
  {Blennow}}, \bibinfo {author} {\bibfnamefont {E.}~\bibnamefont
  {Fernandez-Martinez}}, \bibinfo {author} {\bibfnamefont {T.}~\bibnamefont
  {Ota}},\ and\ \bibinfo {author} {\bibfnamefont {S.}~\bibnamefont
  {Rosauro-Alcaraz}},\ }\bibfield  {title} {\bibinfo {title} {{Physics
  potential of the ESS$\nu$SB}},\ }\href
  {https://doi.org/10.1140/epjc/s10052-020-7761-9} {\bibfield  {journal}
  {\bibinfo  {journal} {Eur. Phys. J. C}\ }\textbf {\bibinfo {volume} {80}},\
  \bibinfo {pages} {190} (\bibinfo {year} {2020})},\ \Eprint
  {https://arxiv.org/abs/1912.04309} {arXiv:1912.04309 [hep-ph]} \BibitemShut
  {NoStop}%
\bibitem [{\citenamefont {Hagiwara}\ \emph {et~al.}(2011)\citenamefont
  {Hagiwara}, \citenamefont {Okamura},\ and\ \citenamefont
  {Senda}}]{Hagiwara:2011kw}%
  \BibitemOpen
  \bibfield  {author} {\bibinfo {author} {\bibfnamefont {K.}~\bibnamefont
  {Hagiwara}}, \bibinfo {author} {\bibfnamefont {N.}~\bibnamefont {Okamura}},\
  and\ \bibinfo {author} {\bibfnamefont {K.-i.}\ \bibnamefont {Senda}},\
  }\bibfield  {title} {\bibinfo {title} {{The earth matter effects in neutrino
  oscillation experiments from Tokai to Kamioka and Korea}},\ }\href
  {https://doi.org/10.1007/JHEP09(2011)082} {\bibfield  {journal} {\bibinfo
  {journal} {JHEP}\ }\textbf {\bibinfo {volume} {09}},\ \bibinfo {pages}
  {082}},\ \Eprint {https://arxiv.org/abs/1107.5857} {arXiv:1107.5857 [hep-ph]}
  \BibitemShut {NoStop}%
\bibitem [{\citenamefont {Roe}(2017)}]{Roe:2017zdw}%
  \BibitemOpen
  \bibfield  {author} {\bibinfo {author} {\bibfnamefont {B.}~\bibnamefont
  {Roe}},\ }\bibfield  {title} {\bibinfo {title} {{Matter density versus
  distance for the neutrino beam from Fermilab to Lead, South Dakota, and
  comparison of oscillations with variable and constant density}},\ }\href
  {https://doi.org/10.1103/PhysRevD.95.113004} {\bibfield  {journal} {\bibinfo
  {journal} {Phys. Rev. D}\ }\textbf {\bibinfo {volume} {95}},\ \bibinfo
  {pages} {113004} (\bibinfo {year} {2017})},\ \Eprint
  {https://arxiv.org/abs/1707.02322} {arXiv:1707.02322 [hep-ex]} \BibitemShut
  {NoStop}%
\bibitem [{\citenamefont {Miura}\ \emph {et~al.}(2001)\citenamefont {Miura},
  \citenamefont {Shindou}, \citenamefont {Takasugi},\ and\ \citenamefont
  {Yoshimura}}]{Miura:2001pi}%
  \BibitemOpen
  \bibfield  {author} {\bibinfo {author} {\bibfnamefont {T.}~\bibnamefont
  {Miura}}, \bibinfo {author} {\bibfnamefont {T.}~\bibnamefont {Shindou}},
  \bibinfo {author} {\bibfnamefont {E.}~\bibnamefont {Takasugi}},\ and\
  \bibinfo {author} {\bibfnamefont {M.}~\bibnamefont {Yoshimura}},\ }\bibfield
  {title} {\bibinfo {title} {{The Matter Fluctuation Effect to T Violation at a
  Neutrino Factory}},\ }\href {https://doi.org/10.1103/PhysRevD.64.073017}
  {\bibfield  {journal} {\bibinfo  {journal} {Phys. Rev. D}\ }\textbf {\bibinfo
  {volume} {64}},\ \bibinfo {pages} {073017} (\bibinfo {year} {2001})},\
  \Eprint {https://arxiv.org/abs/hep-ph/0106086} {arXiv:hep-ph/0106086}
  \BibitemShut {NoStop}%
\bibitem [{\citenamefont {Yokomakura}\ \emph {et~al.}(2002)\citenamefont
  {Yokomakura}, \citenamefont {Kimura},\ and\ \citenamefont
  {Takamura}}]{Yokomakura:2002av}%
  \BibitemOpen
  \bibfield  {author} {\bibinfo {author} {\bibfnamefont {H.}~\bibnamefont
  {Yokomakura}}, \bibinfo {author} {\bibfnamefont {K.}~\bibnamefont {Kimura}},\
  and\ \bibinfo {author} {\bibfnamefont {A.}~\bibnamefont {Takamura}},\
  }\bibfield  {title} {\bibinfo {title} {{Overall Feature of CP Dependence for
  Neutrino Oscillation Probability in Arbitrary Matter Profile}},\ }\href
  {https://doi.org/10.1016/S0370-2693(02)02545-5} {\bibfield  {journal}
  {\bibinfo  {journal} {Phys. Lett. B}\ }\textbf {\bibinfo {volume} {544}},\
  \bibinfo {pages} {286} (\bibinfo {year} {2002})},\ \Eprint
  {https://arxiv.org/abs/hep-ph/0207174} {arXiv:hep-ph/0207174} \BibitemShut
  {NoStop}%
\bibitem [{\citenamefont {Fernandez-Mart{\'\i ne}z}\ \emph
  {et~al.}(2007)\citenamefont {Fernandez-Mart{\'\i ne}z}, \citenamefont
  {Gavela}, \citenamefont {Lopez-Pavon},\ and\ \citenamefont
  {Yasuda}}]{FernandezMartinez:2007ms}%
  \BibitemOpen
  \bibfield  {author} {\bibinfo {author} {\bibfnamefont {E.}~\bibnamefont
  {Fernandez-Mart{\'\i ne}z}}, \bibinfo {author} {\bibfnamefont {M.~B.}\
  \bibnamefont {Gavela}}, \bibinfo {author} {\bibfnamefont {J.}~\bibnamefont
  {Lopez-Pavon}},\ and\ \bibinfo {author} {\bibfnamefont {O.}~\bibnamefont
  {Yasuda}},\ }\bibfield  {title} {\bibinfo {title} {{Cp-Violation from
  Non-Unitary Leptonic Mixing}},\ }\href
  {https://doi.org/10.1016/j.physletb.2007.03.069} {\bibfield  {journal}
  {\bibinfo  {journal} {Phys. Lett. B}\ }\textbf {\bibinfo {volume} {649}},\
  \bibinfo {pages} {427} (\bibinfo {year} {2007})},\ \Eprint
  {https://arxiv.org/abs/hep-ph/0703098} {arXiv:hep-ph/0703098} \BibitemShut
  {NoStop}%
\bibitem [{\citenamefont {Escrihuela}\ \emph {et~al.}(2017)\citenamefont
  {Escrihuela}, \citenamefont {Forero}, \citenamefont {Miranda}, \citenamefont
  {T\'ortola},\ and\ \citenamefont {Valle}}]{Escrihuela:2016ube}%
  \BibitemOpen
  \bibfield  {author} {\bibinfo {author} {\bibfnamefont {F.~J.}\ \bibnamefont
  {Escrihuela}}, \bibinfo {author} {\bibfnamefont {D.~V.}\ \bibnamefont
  {Forero}}, \bibinfo {author} {\bibfnamefont {O.~G.}\ \bibnamefont {Miranda}},
  \bibinfo {author} {\bibfnamefont {M.}~\bibnamefont {T\'ortola}},\ and\
  \bibinfo {author} {\bibfnamefont {J.~W.~F.}\ \bibnamefont {Valle}},\
  }\bibfield  {title} {\bibinfo {title} {{Probing CP violation with non-unitary
  mixing in long-baseline neutrino oscillation experiments: DUNE as a case
  study}},\ }\href {https://doi.org/10.1088/1367-2630/aa79ec} {\bibfield
  {journal} {\bibinfo  {journal} {New J. Phys.}\ }\textbf {\bibinfo {volume}
  {19}},\ \bibinfo {pages} {093005} (\bibinfo {year} {2017})},\ \Eprint
  {https://arxiv.org/abs/1612.07377} {arXiv:1612.07377 [hep-ph]} \BibitemShut
  {NoStop}%
\bibitem [{\citenamefont {Ge}\ and\ \citenamefont
  {Smirnov}(2016)}]{Ge:2016dlx}%
  \BibitemOpen
  \bibfield  {author} {\bibinfo {author} {\bibfnamefont {S.-F.}\ \bibnamefont
  {Ge}}\ and\ \bibinfo {author} {\bibfnamefont {A.~Y.}\ \bibnamefont
  {Smirnov}},\ }\bibfield  {title} {\bibinfo {title} {{Non-Standard
  Interactions and the CP Phase Measurements in Neutrino Oscillations at Low
  Energies}},\ }\href {https://doi.org/10.1007/JHEP10(2016)138} {\bibfield
  {journal} {\bibinfo  {journal} {JHEP}\ }\textbf {\bibinfo {volume} {10}},\
  \bibinfo {pages} {138}},\ \Eprint {https://arxiv.org/abs/1607.08513}
  {arXiv:1607.08513 [hep-ph]} \BibitemShut {NoStop}%
\bibitem [{\citenamefont {de~Gouv\^ea}\ and\ \citenamefont
  {Kelly}(2016)}]{deGouvea:2015ndi}%
  \BibitemOpen
  \bibfield  {author} {\bibinfo {author} {\bibfnamefont {A.}~\bibnamefont
  {de~Gouv\^ea}}\ and\ \bibinfo {author} {\bibfnamefont {K.~J.}\ \bibnamefont
  {Kelly}},\ }\bibfield  {title} {\bibinfo {title} {{Non-standard Neutrino
  Interactions at DUNE}},\ }\href
  {https://doi.org/10.1016/j.nuclphysb.2016.03.013} {\bibfield  {journal}
  {\bibinfo  {journal} {Nucl. Phys. B}\ }\textbf {\bibinfo {volume} {908}},\
  \bibinfo {pages} {318} (\bibinfo {year} {2016})},\ \Eprint
  {https://arxiv.org/abs/1511.05562} {arXiv:1511.05562 [hep-ph]} \BibitemShut
  {NoStop}%
\bibitem [{\citenamefont {Coloma}\ \emph {et~al.}(2020)\citenamefont {Coloma},
  \citenamefont {Esteban}, \citenamefont {Gonzalez-Garcia},\ and\ \citenamefont
  {Maltoni}}]{Coloma:2019mbs}%
  \BibitemOpen
  \bibfield  {author} {\bibinfo {author} {\bibfnamefont {P.}~\bibnamefont
  {Coloma}}, \bibinfo {author} {\bibfnamefont {I.}~\bibnamefont {Esteban}},
  \bibinfo {author} {\bibfnamefont {M.~C.}\ \bibnamefont {Gonzalez-Garcia}},\
  and\ \bibinfo {author} {\bibfnamefont {M.}~\bibnamefont {Maltoni}},\
  }\bibfield  {title} {\bibinfo {title} {{Improved Global Fit to Non-Standard
  Neutrino Interactions Using Coherent Energy and Timing Data}},\ }\href
  {https://doi.org/10.1007/JHEP02(2020)023} {\bibfield  {journal} {\bibinfo
  {journal} {JHEP}\ }\textbf {\bibinfo {volume} {02}},\ \bibinfo {pages}
  {023}},\ \bibinfo {note} {[Addendum: JHEP 12, 071 (2020)]},\ \Eprint
  {https://arxiv.org/abs/1911.09109} {arXiv:1911.09109 [hep-ph]} \BibitemShut
  {NoStop}%
\bibitem [{\citenamefont {Denton}\ \emph {et~al.}(2021)\citenamefont {Denton},
  \citenamefont {Gehrlein},\ and\ \citenamefont {Pestes}}]{Denton:2020uda}%
  \BibitemOpen
  \bibfield  {author} {\bibinfo {author} {\bibfnamefont {P.~B.}\ \bibnamefont
  {Denton}}, \bibinfo {author} {\bibfnamefont {J.}~\bibnamefont {Gehrlein}},\
  and\ \bibinfo {author} {\bibfnamefont {R.}~\bibnamefont {Pestes}},\
  }\bibfield  {title} {\bibinfo {title} {{$CP$ -Violating Neutrino Nonstandard
  Interactions in Long-Baseline-Accelerator Data}},\ }\href
  {https://doi.org/10.1103/PhysRevLett.126.051801} {\bibfield  {journal}
  {\bibinfo  {journal} {Phys. Rev. Lett.}\ }\textbf {\bibinfo {volume} {126}},\
  \bibinfo {pages} {051801} (\bibinfo {year} {2021})},\ \Eprint
  {https://arxiv.org/abs/2008.01110} {arXiv:2008.01110 [hep-ph]} \BibitemShut
  {NoStop}%
\bibitem [{\citenamefont {Kopp}\ \emph {et~al.}(2013)\citenamefont {Kopp},
  \citenamefont {Machado}, \citenamefont {Maltoni},\ and\ \citenamefont
  {Schwetz}}]{Kopp:2013vaa}%
  \BibitemOpen
  \bibfield  {author} {\bibinfo {author} {\bibfnamefont {J.}~\bibnamefont
  {Kopp}}, \bibinfo {author} {\bibfnamefont {P.~A.~N.}\ \bibnamefont
  {Machado}}, \bibinfo {author} {\bibfnamefont {M.}~\bibnamefont {Maltoni}},\
  and\ \bibinfo {author} {\bibfnamefont {T.}~\bibnamefont {Schwetz}},\
  }\bibfield  {title} {\bibinfo {title} {{Sterile Neutrino Oscillations: the
  Global Picture}},\ }\href {https://doi.org/10.1007/JHEP05(2013)050}
  {\bibfield  {journal} {\bibinfo  {journal} {JHEP}\ }\textbf {\bibinfo
  {volume} {05}},\ \bibinfo {pages} {050}},\ \Eprint
  {https://arxiv.org/abs/1303.3011} {arXiv:1303.3011 [hep-ph]} \BibitemShut
  {NoStop}%
\bibitem [{\citenamefont {Gandhi}\ \emph {et~al.}(2015)\citenamefont {Gandhi},
  \citenamefont {Kayser}, \citenamefont {Masud},\ and\ \citenamefont
  {Prakash}}]{Gandhi:2015xza}%
  \BibitemOpen
  \bibfield  {author} {\bibinfo {author} {\bibfnamefont {R.}~\bibnamefont
  {Gandhi}}, \bibinfo {author} {\bibfnamefont {B.}~\bibnamefont {Kayser}},
  \bibinfo {author} {\bibfnamefont {M.}~\bibnamefont {Masud}},\ and\ \bibinfo
  {author} {\bibfnamefont {S.}~\bibnamefont {Prakash}},\ }\bibfield  {title}
  {\bibinfo {title} {{The Impact of Sterile Neutrinos on CP Measurements at
  Long Baselines}},\ }\href {https://doi.org/10.1007/JHEP11(2015)039}
  {\bibfield  {journal} {\bibinfo  {journal} {JHEP}\ }\textbf {\bibinfo
  {volume} {11}},\ \bibinfo {pages} {039}},\ \Eprint
  {https://arxiv.org/abs/1508.06275} {arXiv:1508.06275 [hep-ph]} \BibitemShut
  {NoStop}%
\bibitem [{\citenamefont {Palazzo}(2016)}]{Palazzo:2015gja}%
  \BibitemOpen
  \bibfield  {author} {\bibinfo {author} {\bibfnamefont {A.}~\bibnamefont
  {Palazzo}},\ }\bibfield  {title} {\bibinfo {title} {{3-flavor and 4-flavor
  implications of the latest T2K and NO$\nu$A electron (anti-)neutrino
  appearance results}},\ }\href
  {https://doi.org/10.1016/j.physletb.2016.03.061} {\bibfield  {journal}
  {\bibinfo  {journal} {Phys. Lett. B}\ }\textbf {\bibinfo {volume} {757}},\
  \bibinfo {pages} {142} (\bibinfo {year} {2016})},\ \Eprint
  {https://arxiv.org/abs/1509.03148} {arXiv:1509.03148 [hep-ph]} \BibitemShut
  {NoStop}%
\bibitem [{\citenamefont {Berryman}\ \emph {et~al.}(2015)\citenamefont
  {Berryman}, \citenamefont {de~Gouv\^ea}, \citenamefont {Kelly},\ and\
  \citenamefont {Kobach}}]{Berryman:2015nua}%
  \BibitemOpen
  \bibfield  {author} {\bibinfo {author} {\bibfnamefont {J.~M.}\ \bibnamefont
  {Berryman}}, \bibinfo {author} {\bibfnamefont {A.}~\bibnamefont
  {de~Gouv\^ea}}, \bibinfo {author} {\bibfnamefont {K.~J.}\ \bibnamefont
  {Kelly}},\ and\ \bibinfo {author} {\bibfnamefont {A.}~\bibnamefont
  {Kobach}},\ }\bibfield  {title} {\bibinfo {title} {{Sterile neutrino at the
  Deep Underground Neutrino Experiment}},\ }\href
  {https://doi.org/10.1103/PhysRevD.92.073012} {\bibfield  {journal} {\bibinfo
  {journal} {Phys. Rev. D}\ }\textbf {\bibinfo {volume} {92}},\ \bibinfo
  {pages} {073012} (\bibinfo {year} {2015})},\ \Eprint
  {https://arxiv.org/abs/1507.03986} {arXiv:1507.03986 [hep-ph]} \BibitemShut
  {NoStop}%
\bibitem [{\citenamefont {Antusch}\ \emph {et~al.}(2006)\citenamefont
  {Antusch}, \citenamefont {Biggio}, \citenamefont {Fernandez-Mart{\'\i ne}z},
  \citenamefont {Gavela},\ and\ \citenamefont {Lopez-Pavon}}]{Antusch:2006vwa}%
  \BibitemOpen
  \bibfield  {author} {\bibinfo {author} {\bibfnamefont {S.}~\bibnamefont
  {Antusch}}, \bibinfo {author} {\bibfnamefont {C.}~\bibnamefont {Biggio}},
  \bibinfo {author} {\bibfnamefont {E.}~\bibnamefont {Fernandez-Mart{\'\i
  ne}z}}, \bibinfo {author} {\bibfnamefont {M.~B.}\ \bibnamefont {Gavela}},\
  and\ \bibinfo {author} {\bibfnamefont {J.}~\bibnamefont {Lopez-Pavon}},\
  }\bibfield  {title} {\bibinfo {title} {{Unitarity of the Leptonic Mixing
  Matrix}},\ }\href {https://doi.org/10.1088/1126-6708/2006/10/084} {\bibfield
  {journal} {\bibinfo  {journal} {JHEP}\ }\textbf {\bibinfo {volume} {10}},\
  \bibinfo {pages} {084}},\ \Eprint {https://arxiv.org/abs/hep-ph/0607020}
  {arXiv:hep-ph/0607020} \BibitemShut {NoStop}%
\bibitem [{\citenamefont {Escrihuela}\ \emph {et~al.}(2015)\citenamefont
  {Escrihuela}, \citenamefont {Forero}, \citenamefont {Miranda}, \citenamefont
  {Tortola},\ and\ \citenamefont {Valle}}]{Escrihuela:2015wra}%
  \BibitemOpen
  \bibfield  {author} {\bibinfo {author} {\bibfnamefont {F.~J.}\ \bibnamefont
  {Escrihuela}}, \bibinfo {author} {\bibfnamefont {D.~V.}\ \bibnamefont
  {Forero}}, \bibinfo {author} {\bibfnamefont {O.~G.}\ \bibnamefont {Miranda}},
  \bibinfo {author} {\bibfnamefont {M.}~\bibnamefont {Tortola}},\ and\ \bibinfo
  {author} {\bibfnamefont {J.~W.~F.}\ \bibnamefont {Valle}},\ }\bibfield
  {title} {\bibinfo {title} {{On the description of nonunitary neutrino
  mixing}},\ }\href {https://doi.org/10.1103/PhysRevD.92.053009} {\bibfield
  {journal} {\bibinfo  {journal} {Phys. Rev. D}\ }\textbf {\bibinfo {volume}
  {92}},\ \bibinfo {pages} {053009} (\bibinfo {year} {2015})},\ \bibinfo {note}
  {[Erratum: Phys.Rev.D 93, 119905 (2016)]},\ \Eprint
  {https://arxiv.org/abs/1503.08879} {arXiv:1503.08879 [hep-ph]} \BibitemShut
  {NoStop}%
\bibitem [{\citenamefont {Fong}\ \emph {et~al.}(2019)\citenamefont {Fong},
  \citenamefont {Minakata},\ and\ \citenamefont {Nunokawa}}]{Fong:2017gke}%
  \BibitemOpen
  \bibfield  {author} {\bibinfo {author} {\bibfnamefont {C.~S.}\ \bibnamefont
  {Fong}}, \bibinfo {author} {\bibfnamefont {H.}~\bibnamefont {Minakata}},\
  and\ \bibinfo {author} {\bibfnamefont {H.}~\bibnamefont {Nunokawa}},\
  }\bibfield  {title} {\bibinfo {title} {{Non-Unitary Evolution of Neutrinos in
  Matter and the Leptonic Unitarity Test}},\ }\href
  {https://doi.org/10.1007/JHEP02(2019)015} {\bibfield  {journal} {\bibinfo
  {journal} {JHEP}\ }\textbf {\bibinfo {volume} {02}},\ \bibinfo {pages}
  {015}},\ \Eprint {https://arxiv.org/abs/1712.02798} {arXiv:1712.02798
  [hep-ph]} \BibitemShut {NoStop}%
\bibitem [{\citenamefont {Blennow}\ \emph {et~al.}(2017)\citenamefont
  {Blennow}, \citenamefont {Coloma}, \citenamefont {Fernandez-Mart{\'\i ne}z},
  \citenamefont {Hernandez-Garcia},\ and\ \citenamefont
  {Lopez-Pavon}}]{Blennow:2016jkn}%
  \BibitemOpen
  \bibfield  {author} {\bibinfo {author} {\bibfnamefont {M.}~\bibnamefont
  {Blennow}}, \bibinfo {author} {\bibfnamefont {P.}~\bibnamefont {Coloma}},
  \bibinfo {author} {\bibfnamefont {E.}~\bibnamefont {Fernandez-Mart{\'\i
  ne}z}}, \bibinfo {author} {\bibfnamefont {J.}~\bibnamefont
  {Hernandez-Garcia}},\ and\ \bibinfo {author} {\bibfnamefont {J.}~\bibnamefont
  {Lopez-Pavon}},\ }\bibfield  {title} {\bibinfo {title} {{Non-Unitarity,
  Sterile Neutrinos, and Non-Standard Neutrino Interactions}},\ }\href
  {https://doi.org/10.1007/JHEP04(2017)153} {\bibfield  {journal} {\bibinfo
  {journal} {JHEP}\ }\textbf {\bibinfo {volume} {04}},\ \bibinfo {pages}
  {153}},\ \Eprint {https://arxiv.org/abs/1609.08637} {arXiv:1609.08637
  [hep-ph]} \BibitemShut {NoStop}%
\bibitem [{\citenamefont {Sakurai}\ and\ \citenamefont
  {Napolitano}(2017)}]{sakurai}%
  \BibitemOpen
  \bibfield  {author} {\bibinfo {author} {\bibfnamefont {J.}~\bibnamefont
  {Sakurai}}\ and\ \bibinfo {author} {\bibfnamefont {J.}~\bibnamefont
  {Napolitano}},\ }\href@noop {} {\emph {\bibinfo {title} {Modern Quantum
  Mechanics}}}\ (\bibinfo  {publisher} {Cambridge University Press},\ \bibinfo
  {year} {2017})\BibitemShut {NoStop}%
\bibitem [{\citenamefont {Falkowski}\ \emph {et~al.}(2020)\citenamefont
  {Falkowski}, \citenamefont {Gonz\'alez-Alonso},\ and\ \citenamefont
  {Tabrizi}}]{Falkowski:2019kfn}%
  \BibitemOpen
  \bibfield  {author} {\bibinfo {author} {\bibfnamefont {A.}~\bibnamefont
  {Falkowski}}, \bibinfo {author} {\bibfnamefont {M.}~\bibnamefont
  {Gonz\'alez-Alonso}},\ and\ \bibinfo {author} {\bibfnamefont
  {Z.}~\bibnamefont {Tabrizi}},\ }\bibfield  {title} {\bibinfo {title}
  {{Consistent QFT description of non-standard neutrino interactions}},\ }\href
  {https://doi.org/10.1007/JHEP11(2020)048} {\bibfield  {journal} {\bibinfo
  {journal} {JHEP}\ }\textbf {\bibinfo {volume} {11}},\ \bibinfo {pages}
  {048}},\ \Eprint {https://arxiv.org/abs/1910.02971} {arXiv:1910.02971
  [hep-ph]} \BibitemShut {NoStop}%
\bibitem [{\citenamefont {Bischer}\ and\ \citenamefont
  {Rodejohann}(2019)}]{Bischer:2019ttk}%
  \BibitemOpen
  \bibfield  {author} {\bibinfo {author} {\bibfnamefont {I.}~\bibnamefont
  {Bischer}}\ and\ \bibinfo {author} {\bibfnamefont {W.}~\bibnamefont
  {Rodejohann}},\ }\bibfield  {title} {\bibinfo {title} {{General neutrino
  interactions from an effective field theory perspective}},\ }\href
  {https://doi.org/10.1016/j.nuclphysb.2019.114746} {\bibfield  {journal}
  {\bibinfo  {journal} {Nucl. Phys. B}\ }\textbf {\bibinfo {volume} {947}},\
  \bibinfo {pages} {114746} (\bibinfo {year} {2019})},\ \Eprint
  {https://arxiv.org/abs/1905.08699} {arXiv:1905.08699 [hep-ph]} \BibitemShut
  {NoStop}%
\end{thebibliography}%

\end{document}